\documentclass[namedreferences]{solarphysics}
\usepackage[optionalrh,solaenum]{spr-sola-addons} 
\usepackage{graphicx}                    
\usepackage{amssymb}                    
\usepackage{color}                       
\usepackage{url}                         
\usepackage{lscape}
\usepackage{hangcaption}

\begin{document}

\begin{article}

\begin{opening}

\title{Kinematic Properties of Slow ICMEs and an Interpretation of a Modified Drag Equation for Fast and Moderate ICMEs}

%
\author{T.~\surname{Iju}$^{1}$\sep
        M.~\surname{Tokumaru}$^{1}$\sep
        K.~\surname{Fujiki}$^{1}$      
       }

%

%
  \institute{$^{1}$ Solar-Terrestrial Environment Laboratory, Nagoya University, Furo-cho, Chikusa-ku, Nagoya, Aichi 464-8601, Japan \\
  email: \url{tomoya@stelab.nagoya-u.ac.jp} \\ 
             }

\begin{abstract}
We report kinematic properties of slow interplanetary coronal mass ejections 
(ICMEs) identified by SOHO/LASCO, 
interplanetary scintillation, and \textit{in-situ} observations, and propose a modified equation for the ICME 
motion. We identify seven ICMEs between 2010 and 2011, and examine them with 39 events 
reported in our previous work.
We examine 15 fast ($V_{\mathrm{SOHO}} - V_\mathrm{bg} > 500$ ${\mathrm{km~s^{-1}}}$), 
25 moderate ($0$ $\mathrm{km~s^{-1}}$ $\le V_{\mathrm{SOHO}} - V_\mathrm{bg} \le 500$ ${\mathrm{km~s^{-1}}}$), 
and 6 slow ($V_{\mathrm{SOHO}} - V_\mathrm{bg} <$ $0$ ${\mathrm{km~s^{-1}}}$) ICMEs, where ${V_{\mathrm{SOHO}}}$ and ${V_{\mathrm{bg}}}$ are
the initial speed of ICMEs and the speed of the background solar wind, respectively. 
For slow ICMEs, we found the following results: i) They accelerate toward the speed of the background solar wind during 
their propagation, and reach their final speed by ${0.34 \pm 0.03}$ AU. ii) The acceleration ends when they reach 
$479 \pm 126$ ${\mathrm{km~s^{-1}}}$; 
this is close to the typical speed of the solar wind during the period of this study. 
iii) When ${{\gamma}_\mathrm{1}}$ and ${{\gamma}_\mathrm{2}}$ are assumed to be constants, 
a quadratic equation for the acceleration $a = -{\gamma}_{\mathrm{2}}(V - V_\mathrm{bg})|V - V_\mathrm{bg}|$ is 
more appropriate than a linear one $a = -{\gamma}_{\mathrm{1}}(V - V_\mathrm{bg})$, 
where $V$ is the propagation speed of ICMEs, while the latter gives a smaller ${\chi}^{2}$ value than the former. 
For the motion of the fast and moderate ICMEs, we found 
a modified drag equation $a = -ð2.07 {\times} 10^{-12}(V - V_\mathrm{bg})|V - V_\mathrm{bg}|-ð4.84 {\times}10^{-6}(V - V_\mathrm{bg})$. 
From the viewpoint of fluid dynamics, we interpret this equation as indicating that 
ICMEs with ${0}$ ${\mathrm{km~s^{-1}}}$ $\le V - V_\mathrm{bg} \le 2300$ ${\mathrm{km~s^{-1}}}$ are controlled mainly by 
the hydrodynamic Stokes drag force, while the aerodynamic drag force is a predominant factor for the 
propagation of ICME with $V - V_\mathrm{bg} > 2300$ ${\mathrm{km~s^{-1}}}$. 
\end{abstract}

%

\end{opening}

 \section{Introduction}
         \label{introduction} 

Coronal mass ejections (CMEs) are transient events in which a large amount of plasma and 
magnetic field are expelled from the Sun into the interplanetary space with a wide 
range of speed. Interplanetary coronal mass ejections (ICMEs) are defined as CMEs propagating 
far from the Sun \cite{ICMEs}. Some of them reach the Earth and sometimes cause severe geomagnetic storms 
\cite{Gosling1991,Brueckner1998,Cane2000}. 
Therefore, understanding of ICME propagation is very important for space weather forecasting. 
It is known that the range of ICME speeds in the near-Earth region is narrower than that in 
the near-Sun region by space-borne coronagraphs and near-Earth \textit{in-situ} observations 
(\textit{e.g.} \opencite{Lindsay1999}; \opencite{Gopalswamy2000}, \citeyear{Gopalswamy2001}). 
On the basis of this fact, 
many investigators expect that ICMEs undergo an interaction with the ambient interplanetary 
medium. \inlinecite{Vrsnak2002} proposed a model for the motion of ICMEs in which 
the interaction with the solar wind is simply expressed by the following equation for the acceleration:
\begin{equation}
  \label{eq.linear}
  a = - {\gamma}_{\mathrm{1}}(V - V_\mathrm{bg}), 
\end{equation}
where ${{\gamma}_{\mathrm{1}}}$ is a function of distance, and $V$ and $V_\mathrm{bg}$ are the speeds of ICMEs and of the 
background solar wind, respectively. They also compared their model with the drag acceleration of the following form:
\begin{equation}
  \label{eq.quadratic}
  a = - {\gamma}_{\mathrm{2}}(V - V_\mathrm{bg})|V - V_\mathrm{bg}|, 
\end{equation}
where ${{\gamma}_\mathrm{2}}$ is another function of distance; this expression is known as 
the aerodynamic drag force (\textit{e.g.} \opencite{Chen1996}; \opencite{Cargill2004}; 
\opencite{Vrsnak2010}). 
Although the motion of ICMEs is also affected by the Lorentz and gravity forces in the near-Sun region, 
it is expected that both forces become negligible at a large distance \cite{Chen1996}. 
Therefore, they take only the effect of drag force into account. \citeauthor{Borgazzi2009} 
(\citeyear{Borgazzi2008}, \citeyear{Borgazzi2009}) studied the dynamics of ICMEs in the solar wind 
using the hydrodynamic theory. They introduced two kinds of drag force depending on $V - V_\mathrm{bg}$ 
(a laminar drag) and $(V - V_\mathrm{bg})^2$ (a turbulent drag) to the equation of motion. 

 Drag force models have been tested by comparing with ICME observations. \inlinecite{Reiner2003} 
examined the speed profile of a CME obtained by measurements of type-II radio bursts. 
\inlinecite{Tappin2006} studied the propagation of a CME that occurred on 5 April 2003 using 
observations by the \textit{Large Angle and Spectrometric Coronagraph} (LASCO; \opencite{Brueckner1995}) onboard the 
\textit{Solar and Heliosphere Observatory} (SOHO) spacecraft, \textit{Solar Mass Ejection Imager} (SMEI)
onboard the \textit{Coriolis} satellite, and the \textit{Ulysses} interplanetary probe. 
\inlinecite{Manoharan2006} examined radial evolutions of 30 CMEs observed by SOHO/LASCO, 
\textit{Advanced Composition Explorer} (ACE; \opencite{Stone1998}), and the Ooty radio telescope between 
1998 and 2004. \inlinecite{Maloney2010} derived three-dimensional kinematics of three ICMEs 
detected between 2008 and 2009 using the \textit{Sun Earth Connection Coronal and Heliospheric Investigation} (SECCHI) instruments 
onboard the \textit{Solar-Terrestrial Relations Observatory} (STEREO) A and B spacecraft. 
\inlinecite{Temmer2011} also studied the influence of the solar wind on the propagation of 
three ICMEs using the same spacecraft. \inlinecite{Lara2011} investigated the velocity profile of an ICME 
from the Sun to 5.3 AU using data from SOHO/LASCO, ACE, and \textit{Ulysses}, and then estimated the drag coefficient 
and the kinematic viscosity for the ICME--solar wind interaction on the basis of fluid dynamics.

 We assume that the radial motion of ICMEs is governed by the drag force(s) due to interaction with 
the background solar wind, and that the magnitude of the force is proportional to the difference in 
speed between the ICME and the solar wind. Our assumption should be tested using data obtained by interplanetary observations. 
We take advantage of the interplanetary scintillation (IPS; \opencite{Hewish1964}) observations to 
determine the speeds and accelerations of ICMEs. Our IPS observations have been carried out since 
early 1980s using the 327 MHz radio-telescope system of the Solar-Terrestrial Environment Laboratory 
(STEL), Nagoya University \cite{Kojima1990}. These observations allow us to probe into the inner heliosphere with 
a cadence of 24 h, and therefore are suitable to collecting global data on ICMEs. 

 In our previous study (\opencite{Iju2013}; referred to as Paper I), we detected 39 ICMEs using the IPS 
observations by the Kiso IPS antenna \cite{Asai1995} during 1997\,--\,2009. 
Using the values of the initial speed ($V_\mathrm{SOHO}$) and $V_\mathrm{bg}$, 
we classified them into three types: fast ($V_{\mathrm{SOHO}} - V_\mathrm{bg} > 500$ $\mathrm{km~s^{-1}}$), 
moderate ($0~{\mathrm{km~s^{-1}}} \le V_{\mathrm{SOHO}} - V_\mathrm{bg} \le 500$ ${\mathrm{km~s^{-1}}}$), and slow 
($V_{\mathrm{SOHO}} - V_\mathrm{bg} < 0$ $\mathrm{km~s^{-1}}$), and then examined their kinematic properties. 
From this examination, we found that fast and moderate ICMEs decelerate, while slow ones accelerate, and their radial speeds 
converge toward the speed of the solar wind as the distance increases. We also found that Equation (\ref{eq.linear}) is more 
appropriate than (\ref{eq.quadratic}) to describe the kinematics of ICMEs moving faster than the solar wind.

 In the current study, we add new ICMEs identified between 2010 and 2011 to our list, and then examine their kinematics again 
on the assumption that ICMEs are controlled by the drag force(s) only. Earlier observational studies were mainly 
on the propagation of fast ICMEs, although the propagation of slow ICMEs was also studied 
(\textit{e.g.} \opencite{Shanmugaraju2009}; \opencite{Byrne2010}; \opencite{Maloney2010}; \opencite{Lynch2010}; 
\opencite{Temmer2011}; \opencite{Shen2011}; \opencite{Rollett2012}; \opencite{Vrsnak2013}). 
These earlier studies presented mainly case studies of slow ICMEs. However, 
understanding the general properties of their propagation requires a statistical study. Hence, in this article we focus on 
the kinematics of slow ICMEs, and determine their general properties by statistical analysis. 
We also examine fast and moderate ICMEs in further detail. Although we showed a simple equation for their motion in 
Paper I, we will provide a modified one and its physical implications in this article.

 The outline of this article is as follows. 
Section~\ref{data and method} describes IPS observations, methods for event 
identification, and estimating ICME speeds and accelerations. 
Section~\ref{results} provides the speed profiles of ICMEs and the analyses of the propagation properties. 
Section~\ref{discussion} discusses the propagation of slow ICMEs, a modified drag equation for fast and moderate ICMEs, and the 
estimated viscosity for the ICME--solar wind interaction. 
Section~\ref{conclusion} summarizes the main conclusions of our study.

\section{Data and Analysis Method} 
      \label{data and method}      

The solar wind disturbance factor, the so called ``\textit{g}-value'' 
\cite{Gapper1982}, is derived from IPS observations, and represents the relative level of density 
fluctuation integrated along the line-of-sight (LOS) from an observed radio source to a telescope.
When dense plasma passes across the LOS, the \textit{g}-value becomes 
greater than unity, while that is about unity for the quiet solar wind. 
In the current study, we use \textit{g}-value data obtained between 2010 and 2011. The measurement of 
\textit{g}-value has been carried out using the Solar Wind Imaging Facility (SWIFT; \opencite{Tokumaru2011}) since 2010.
From an examination of these data, we found 260 disturbance days between 2010 and 2011, and made a list of them.

These disturbance days should be compared with CME/ICME pairs identified using SOHO/LASCO and \textit{in-situ} observations. 
In this examination, because there was no list of CME/ICME pairs between 2010 and 2011, 
we identified them ourselves using the SOHO/LASCO CME catalog (\opencite{Yashiro2004}; \opencite{Gopalswamy2009}; 
available at \url{cdaw.gsfc.nasa.gov/CME_list/}), 1 and 2 h averaged data of solar-wind 
charge states obtained by the \textit{Solar Wind Ion Composition Spectrometer} (SWICS; \opencite{Gloeckler1998}) onboard ACE 
(available at \url{www.srl.caltech.edu/ACE/ASC/}), 
and the criteria of ICME identification introduced by \inlinecite{Richardson2010}. 
According to the above paper, the mean Fe charge ${\langle \mathrm{Q}_\mathrm{Fe} \rangle}$ and O\,{\sc viii}\,/\,O\,{\sc vii} 
ratio are enhanced during the passage of an ICME. Hence, we define the detection of a near-Earth ICME as 
the enhancement in the charge state observed by ACE/SWICS within five days after the appearance of a major Earth-side CME 
in the SOHO/LASCO-C2 field-of-view (FOV). 
The start and end times of an ICME event correspond to those of the charge-state enhancement, respectively. 
Using the above method, we made a list of CME/ICMEs found between 2010 and 2011.

We compared the list of disturbance days with that of CME/ICMEs by assuming that an ICME causes 
a disturbance day. We then identified seven ICMEs that were detected by SOHO/LASCO, IPS, and 
\textit{in-situ} observations between 2010 and 2011. For them, we calculated the average reference distances 
(${R_\mathrm{1}}$ and ${R_\mathrm{2}}$), the average radial speeds (${V_\mathrm{1}}$ and ${V_\mathrm{2}}$), and 
accelerations (${a_\mathrm{1}}$ and ${a_\mathrm{2}}$). We also estimated the transit speed (${V_\mathrm{Tr}}$) using the appearance time in 
the SOHO/LASCO-C2 FOV (${T_{\mathrm{SOHO}}}$) and the detection time at 1 AU (${T_{\mathrm{Earth}}}$). 
The initial speed of the associated CMEs (${V_{\mathrm{SOHO}}}$) was estimated from their speed measured in 
the plane of the sky by SOHO/LASCO (${V_{\mathrm{POS}}}$). The radial speed of near-Earth ICMEs (${V_{\mathrm{Earth}}}$) 
is equivalent to the speed of plasma flow during the charge-state enhancement derived from \textit{in-situ} measurements. 
We note that ${V_{\mathrm{SOHO}}}$ and ${V_{\mathrm{Earth}}}$ represent the average values in the near-Sun and near-Earth regions. 
Linkewise, ${V_\mathrm{1}}$ and ${a_\mathrm{1}}$ are averages in the SOHO--IPS region (from 0.1 to ${\approx}$ 0.6 AU), 
and ${V_\mathrm{2}}$ and ${a_\mathrm{2}}$ are averages in the IPS--Earth region (from ${\approx}$ 0.6 to 1 AU). 
To determine the speed of the background solar wind ${V_\mathrm{bg}}$, we used the OMNI dataset through 
OMNIWeb Plus (\url{omniweb.gsfc.nasa.gov/}). 
Using the value of ${V_{\mathrm{SOHO}}}$ and ${V_\mathrm{bg}}$, we have classified seven ICMEs into 
fast ($V_{\mathrm{SOHO}} - V_\mathrm{bg} > 500$ ${\mathrm{km~s^{-1}}}$), 
moderate ($0$ ${\mathrm{km~s^{-1}}}$ $\le V_{\mathrm{SOHO}} - V_\mathrm{bg} \le 500$ ${\mathrm{km~s^{-1}}}$), and 
slow ($V_{\mathrm{SOHO}} - V_\mathrm{bg} <$ $0$ ${\mathrm{km~s^{-1}}}$) ones. 
In our results, the numbers of fast, moderate, and slow ICMEs are 1, 5, and 1, respectively. 
Detailed methods of calculation for the above properties were presented in Paper I.

\section{Results} 
      \label{results}
     
We list the properties of seven ICMEs detected by SWIFT between 2010 and 2011 in Tables \ref{table1} and \ref{table2} including ${R_\mathrm{0}}$, 
${T_{\mathrm{IPS}}}$, ${\alpha}$, and ${\beta}$ in addition to ${T_{\mathrm{SOHO}}}$, ${V_{\mathrm{POS}}}$, ${V_{\mathrm{SOHO}}}$, ${R_\mathrm{1}}$, 
${V_\mathrm{1}}$, ${a_\mathrm{1}}$, ${R_\mathrm{2}}$, ${V_\mathrm{2}}$, ${a_\mathrm{2}}$, ${T_{\mathrm{Earth}}}$, ${V_{\mathrm{Earth}}}$, 
${V_\mathrm{Tr}}$, and ${V_\mathrm{bg}}$. 
Here, ${T_{\mathrm{IPS}}}$ and ${R_\mathrm{0}}$ are the mean detection time and average radial distance for 
a disturbance detected by IPS observations, respectively; their detailed descriptions were presented in Paper I.
Parameters ${\alpha}$ and ${\beta}$ are defined as
\begin{equation}
  \label{eq.powerlaw}
V = {\beta}R^{\alpha}, 
\end{equation}
where $R$ is the heliocentric distance. On the other hand, the catalog of 39 ICMEs detected by the Kiso IPS antenna during 1997\,--\,2009 
was given in Paper I.
We also provide a list of slow ICMEs extracted from the above catalog in Tables \ref{table3} and \ref{table4}. 
Therefore, we examine 46 ICMEs which consist of 15 fast, 25 moderate, and 6 slow ones identified during 1997\,--\,2011.
In this investigation, we assume that ${V_\mathrm{bg}}$ is constant for heliocentric distances 
ranging from ${\approx}$ 0.1 to 1 AU. This assumption is consistent with the speed profile of the solar wind estimated using coronagraph 
observations \cite{Sheeley1997,Guhathakurta1998}. The constant speed of the solar wind has been verified between ${\approx}$ 0.3 and 1 AU 
by \textit{in-situ} measurements \cite{Schwenn1981}.
     %
     \captionwidth=17.2cm
     \begin{landscape}
      \begin{table}
      \hangcaption{Properties derived from SOHO/LASCO, IPS (SWIFT), and \textit{in-situ} observations for seven ICMEs during 2010\,--\,2011.}
      \label{table1}
      \scalebox{0.90}{
      \begin{tabular}{cccccccccccccccccc}
      \hline
~~~ & \multicolumn{6}{c}{SOHO/LASCO} &  & \multicolumn{10}{c}{IPS} \\ \cline{2-7} \cline{9-18}
~~~ & ~~~~~~~~~~~ & ~~~~ & ~~~ & ~~~ & ~~~ & ~~~ &  & ~~~~~~~~~~~ & \multicolumn{3}{c}{Disturbance} & \multicolumn{6}{c}{SOHO--IPS region} \\
~~~ & Date & Time & ${V_\mathrm{POS}}$ & ${V_\mathrm{SOHO}}$ & CME & PA &  & Date & Time & \multicolumn{2}{c}{${R_\mathrm{0}}$~[AU]} & \multicolumn{2}{c}{$R_\mathrm{1}$~[AU]} & \multicolumn{2}{c}{$V_\mathrm{1}$~[$\mathrm{km~s^{-1}}$]} & \multicolumn{2}{c}{$a_\mathrm{1}$~[$\mathrm{m~s^{-2}}$]} \\
No. & [ddmmmyyyy] & [hhmm] & [${\mathrm{km~s^{-1}}}$] & [${\mathrm{km~s^{-1}}}$] & Type & [deg] &  & [ddmmmyyyy] & [hhmm] & aver. & ${\sigma}$ & aver. & ${\sigma}$ & aver. & ${\sigma}$ & aver. & ${\sigma}$ \\
      \hline
1 & 07~Feb~2010 & 0354 & 421 & ~505 & FH & $-99$ &  & 11~Feb~2010 & 0117 & 0.81 & 0.20 & 0.45 & 0.10 & ~359 & ~92 & $-0.44$ & 0.56 \\
2 & 03~Apr~2010 & 1033 & 668 & ~802 & FH & $-99$ &  & 04~Apr~2010 & 0043 & 0.81 & 0.16 & 0.44 & 0.08 & 1030 & 236 & $-0.77$ & 1.56 \\
3 & 08~Apr~2010 & 0131 & 227 & ~272 & PH & $~76$ &  & 11~Apr~2010 & 0240 & 0.67 & 0.22 & 0.37 & 0.11 & ~374 & 122 & $~0.90$ & 0.84 \\
4 & 24~May~2010 & 1406 & 427 & ~512 & FH & $-99$ &  & 26~May~2010 & 0251 & 0.63 & 0.19 & 0.35 & 0.09 & ~466 & 135 & $-0.51$ & 0.53 \\
5 & 01~Aug~2010 & 1342 & 850 & 1020 & FH & $-99$ &  & 03~Aug~2010 & 0356 & 0.77 & 0.21 & 0.43 & 0.11 & ~563 & 160 & $-1.70$ & 1.67 \\
6 & 12~Nov~2010 & 0836 & 482 & ~578 & PH & $170$ &  & 15~Nov~2010 & 0201 & 0.80 & 0.16 & 0.44 & 0.08 & ~501 & 105 & $-0.56$ & 0.65 \\
7 & 15~Feb~2011 & 0224 & 669 & ~803 & FH & $-99$ &  & 17~Feb~2011 & 0307 & 0.73 & 0.17 & 0.41 & 0.08 & ~615 & 135 & $-1.49$ & 0.87 \\
     \hline
\multicolumn{18}{p{21.4cm}}{
Column: (1) Event number; (2)\,--\,(3) Appearance date [ddmmmyyyy] and time [hhmm] of 
an ICME--associated CME observed by SOHO/LASCO;(4) Speed in the sky plane measured by SOHO/LASCO at a reference distance of 0.08 AU; 
(5) Radial speed estimated using $V_{\mathrm{SOHO}} = 1.20 \times V_{\mathrm{POS}}$; (6) Type of CME (FH, PH, and 
NM mean Full Halo, Partial Halo, and Normal CME, respectively); 
(7) Position angle measured counter-clockwise from solar north in degrees ($- 99$ means Full Halo); (8)\,--\,(9) Observation date [ddmmmyyyy] and 
mean time [hhmm] of IPS disturbance event day ; (10)\,--\,(11) Average and standard errors for the distance of observed disturbance ${R_\mathrm{0}}$; 
(12)\,--\,(13) Average and standard errors for the reference distance ${R_\mathrm{1}}$ in the SOHO--IPS region; (14)\,--\,(15) Average and standard 
errors for the speed ${V_\mathrm{1}}$ in the SOHO--IPS region; (16)\,--\,(17) Average and standard errors for acceleration ${a_\mathrm{1}}$ 
in the SOHO--IPS region.}
      \end{tabular}
      }
      \end{table}
    \end{landscape}
     %
     \captionwidth=17.2cm
     \begin{landscape}
      \begin{table}
      \hangcaption{Properties derived from SOHO/LASCO, IPS (SWIFT), and \textit{in-situ} observations for seven ICMEs, and speeds of the background solar wind during 2010\,--\,2011.}
      \label{table2}
      \scalebox{0.90}{
      \begin{tabular}{ccccccccccccccccc}
      \hline
~~~ & \multicolumn{6}{c}{IPS} &  & \multicolumn{3}{c}{\textit{in~situ}} & \multicolumn{2}{c}{Parameters for} & ~~~ & ~~~ &~\\ \cline{2-7} \cline{9-11}
~~~ & \multicolumn{6}{c}{IPS--Earth region} & & ~~~ & ~~~ & ~~~ & \multicolumn{2}{c}{power-law equation} & \multicolumn{3}{p{3.7cm}}{~~~~~~~~~~~~~~~Background wind} \\
~~~ & \multicolumn{2}{c}{$R_\mathrm{2}$~[AU]} &\multicolumn{2}{c}{$V_\mathrm{2}$~[$\mathrm{km~s^{-1}}$]} & \multicolumn{2}{c}{$a_\mathrm{2}$~[$\mathrm{m~s^{-2}}$]} & & Date & Time & $V_\mathrm{Earth}$ & Index & Coefficient & $V_\mathrm{Tr}$ & \multicolumn{2}{p{2.2cm}}{~$V_\mathrm{bg}$~[$\mathrm{km~s^{-1}}$]} \\
No. & aver. & ${\sigma}$ & aver. & ${\sigma}$ & aver. & ${\sigma}$ &  & [ddmmmyyyy] & [hhmm] & [$\mathrm{km~s^{-1}}$] & ${\alpha}$ & ${\beta}$ & [$\mathrm{km~s^{-1}}$] & aver. & $\sigma$ \\
      \hline
1 & 0.91 & 0.10 & 356 & 376 & $~0.62$ & 2.29 & & 12~Feb~2010 & 0000 & 410 & $-0.113$ & 366.1 & 358 & 339 & 23 \\
2 & 0.90 & 0.07 & 388 & 300 & $~0.01$ & 2.40 & & 05~Apr~2010 & 1600 & 701 & $-0.160$ & 599.7 & 776 & 540 & 90 \\
3 & 0.83 & 0.11 & 640 & 445 & $-1.10$ & 2.89 & & 12~Apr~2010 & 0000 & 419 & ~$0.255$ & 514.2 & 439 & 461 & 74 \\
4 & 0.81 & 0.09 & 359 & 175 & $-0.25$ & 0.68 & & 28~Apr~2010 & 1600 & 373 & $-0.140$ & 370.3 & 424 & 328 & 22 \\
5 & 0.89 & 0.10 & 786 & 692 & $-1.62$ & 7.97 & & 04~Aug~2010 & 1000 & 592 & $-0.169$ & 620.4 & 608 & 461 & 35 \\
6 & 0.90 & 0.08 & 395 & 303 & $~1.46$ & 1.94 & & 16~Nov~2010 & 1800 & 557 & $-0.080$ & 468.7 & 476 & 534 & 72 \\
7 & 0.87 & 0.08 & 466 & 298 & $-0.38$ & 1.78 & & 18~Feb~2011 & 0300 & 506 & $-0.203$ & 487.3 & 572 & 534 & 46 \\
     \hline
\multicolumn{16}{p{19.7cm}}{
Column: (1) Event number [identical with column (1) in Table \ref{table1}]; (2)\,--\,(3) Average and standard errors for the reference distance $R_\mathrm{2}$ in the IPS--Earth region; 
(4)\,--\,(5) Average and standard errors for the speed $V_\mathrm{2}$ in the IPS--Earth region; (6)\,--\,(7) Average and standard errors for the acceleration $a_\mathrm{2}$ in the IPS--Earth region; 
(8)\,--\,(9) Detection date [ddmmmyyyy] and time [hhmm] of a near-Earth ICME by \textit{in-situ} observation at 1 AU; (10) Near-Earth ICME speed measured by \textit{in-situ} observation at 1 AU; 
(11)\,--\,(12); Index $\alpha$ and coefficient $\beta$ for a power-law form of radial speed evolution; (13) 1 AU transit speed derived from the CME appearance and the ICME detection times; 
(14)\,--\,(15) Average and standard errors for the speed of the background wind $V_\mathrm{bg}$ measured by spacecraft.
} & ~\\
      \end{tabular}
      }
      \end{table}
    \end{landscape}
     %
     \captionwidth=17.2cm
     \begin{landscape}
      \begin{table}[!p]
      \hangcaption{Properties derived from SOHO/LASCO, IPS (KIT and SWIFT), and \textit{in-situ} observations for six slow ICMEs during 1997\,--\,2011.}
      \label{table3}
      \scalebox{0.90}{
      \begin{tabular}{ccccccccccccccccccc}
      \hline
~~~ & \multicolumn{6}{c}{SOHO/LASCO} &  & \multicolumn{10}{c}{IPS} \\ \cline{2-7} \cline{9-18}
~~~ & ~~~~ & ~~~~ & ~~~ & ~~~ & ~~~ & ~~ &  & ~~~~ & \multicolumn{2}{c}{Disturbance} & \multicolumn{6}{c}{SOHO--IPS region} \\
~~~ & Date & Time & ${V_\mathrm{POS}}$ & ${V_\mathrm{SOHO}}$ & CME & PA &  & Date & Time & \multicolumn{2}{c}{${R_\mathrm{0}}$~[AU]} & \multicolumn{2}{c}{${R_\mathrm{1}}$~[AU]} & \multicolumn{2}{c}{${V_\mathrm{1}}$~[${\mathrm{km~s^{-1}}}$]} & \multicolumn{2}{c}{${a_\mathrm{1}}$~[${\mathrm{m~s^{-2}}}$]} \\
No. & [ddmmmyyyy] & [hhmm] & [${\mathrm{km~s^{-1}}}$] & [${\mathrm{km~s^{-1}}}$] & Type & [deg] &  & [ddmmmyyyy] & [hhmm] & aver. & ${\sigma}$ & aver. & ${\sigma}$ & aver. & ${\sigma}$ & aver. & ${\sigma}$ \\
      \hline
1 & 13~Apr~1999 & 0330 & 291 & 349 & PH & $228$ &  & 15~Apr~1999 & 0453 & 0.55 & 0.16 & 0.32 & 0.08 & ~456 & 130 & $~0.71$ & 0.55 \\
2 & 06~Aug~2000 & 1830 & 233 & 280 & PH & $105$ &  & 09~Aug~2000 & 0503 & 0.62 & 0.14 & 0.35 & 0.07 & ~432 & ~95 & $~0.68$ & 0.40 \\
3 & 14~Aug~2003 & 2006 & 378 & 454 & FH & $-99$ &  & 17~Aug~2003 & 0409 & 0.68 & 0.12 & 0.38 & 0.06 & ~497 & 106 & $~0.62$ & 0.63 \\
4 & 12~Sep~2008 & 1030 & ~91 & ~91 & NM & $~89$ &  & 14~Sep~2008 & 0449 & 0.58 & 0.10 & 0.33 & 0.05 & ~556 & ~96 & $~2.04$ & 0.34 \\
5 & 29~May~2009 & 0930 & 139 & 139 & NM & $258$ &  & 01~Jun~2009 & 0148 & 0.56 & 0.18 & 0.32 & 0.09 & ~353 & 116 & $~0.72$ & 0.32 \\
6 & 08~Apr~2010 & 0131 & 227 & 272 & PH & $~76$ &  & 11~Apr~2010 & 0240 & 0.67 & 0.22 & 0.37 & 0.11 & ~374 & 122 & $~0.90$ & 0.84 \\
     \hline
\multicolumn{18}{p{21.4cm}}{
Column: (1) Event number; (2)\,--\,(3) Appearance date [ddmmmyyyy] and time [hhmm] of 
an ICME--associated CME observed by SOHO/LASCO;(4) Speed in the sky plane measured by SOHO/LASCO at a reference 
distance of 0.08 AU ; (5) Radial speed estimated using $V_{\mathrm{SOHO}} = 1.20 \times V_{\mathrm{POS}}$; (6) Type of CME (FH, PH, and 
NM mean Full Halo, Partial Halo, and Normal CME, respectively); 
(7) Position angle measured counter-clockwise from solar north in degrees ($- 99$ means Full Halo); (8)\,--\,(9) Observation date [ddmmmyyyy] and 
mean time [hhmm] of IPS disturbance event day ; (10)\,--\,(11) Average and standard errors for the distance of observed disturbance ${R_\mathrm{0}}$; 
(12)\,--\,(13) Average and standard errors for the reference distance ${R_{1}}$ in the SOHO--IPS region; (14)\,--\,(15) Average and standard 
errors for the speed ${V_\mathrm{1}}$ in the SOHO--IPS region; (16)\,--\,(17) Average and standard errors for acceleration ${a_\mathrm{1}}$ 
in the SOHO--IPS region.}
      \end{tabular}
      }
      \end{table}
     \end{landscape}
     %
     \captionwidth=17.2cm
     \begin{landscape}
      \begin{table}
      \hangcaption{Properties derived from SOHO/LASCO, IPS (KIT and SWIFT), and \textit{in-situ} observations for six slow ICMEs, and speeds of background solar wind during 1997--2011.}
      \label{table4}
      \scalebox{0.90}{
      \begin{tabular}{ccccccccccccccccc}
      \hline
~~~ & \multicolumn{6}{c}{IPS} &  & \multicolumn{3}{c}{\textit{in~situ}} & \multicolumn{2}{c}{Parameters for} & ~~~ &  ~~~ &~\\ \cline{2-7} \cline{9-11}
~~~ & \multicolumn{6}{c}{IPS--Earth region} & & ~~~ & ~~~ & ~~~ & \multicolumn{2}{c}{power-law equation} & \multicolumn{3}{p{3.7cm}}{~~~~~~~~~~~~~~~Background wind} \\
~~~ & \multicolumn{2}{c}{$R_\mathrm{2}$~[AU]} &\multicolumn{2}{c}{$V_\mathrm{2}$~[$\mathrm{km~s^{-1}}$]} & \multicolumn{2}{c}{$a_\mathrm{2}$~[$\mathrm{m~s^{-2}}$]} & & Date & Time & $V_\mathrm{Earth}$ & Index & Coefficient & $V_\mathrm{Tr}$ & \multicolumn{2}{p{2.2cm}}{~$V_\mathrm{bg}$~[$\mathrm{km~s^{-1}}$]} \\
No. & aver. & ${\sigma}$ & aver. & ${\sigma}$ & aver. & ${\sigma}$ &  & [ddmmmyyyy] & [hhmm] & [$\mathrm{km~s^{-1}}$] & ${\alpha}$ & ${\beta}$ & [$\mathrm{km~s^{-1}}$] & aver. & $\sigma$ \\
      \hline
1 & 0.78 & 0.08 & 495 & 165 & $-0.49$ & 0.73 & & 16~Apr~1999 & 1800 & 410 & $0.094$ & 465.0 & 480 & 398 &~15 \\
2 & 0.81 & 0.07 & 414 & 149 & $~0.06$ & 0.62 & & 10~Aug~2000 & 1900 & 430 & $0.165$ & 447.8 & 430 & 412 &~36 \\
3 & 0.84 & 0.06 & 666 & 265 & $-1.84$ & 1.70 & & 18~Aug~2003 & 0100 & 450 & $0.068$ & 543.0 & 540 & 534 &~55 \\
4 & 0.79 & 0.05 & 247 & ~58 & $-0.01$ & 0.20 & & 17~Sep~2008 & 0400 & 400 & $0.486$ & 425.8 & 366 & 406 &107 \\
5 & 0.78 & 0.09 & 256 & 103 & $~0.02$ & 0.28 & & 04~Jun~2009 & 0200 & 310 & $0.276$ & 327.7 & 304 & 327 &~26 \\
6 & 0.83 & 0.11 & 640 & 445 & $-1.10$ & 2.89 & & 12~Apr~2010 & 0000 & 419 & $0.255$ & 514.2 & 439 & 461 &~74 \\
     \hline
\multicolumn{16}{p{19.7cm}}{
Column: (1) Event number [identical with column (1) in Table \ref{table1}]; (2)\,--\,(3) Average and standard errors for the reference distance $R_\mathrm{2}$ in the IPS--Earth region; 
(4)\,--\,(5) Average and standard errors for the speed $V_\mathrm{2}$ in the IPS--Earth region; (6)\,--\,(7) Average and standard errors for the acceleration $a_\mathrm{2}$ in the IPS--Earth region; 
(8)\,--\,(9) Detection date [ddmmmyyyy] and time [hhmm] of a near-Earth ICME by \textit{in-situ} observation at 1 AU; (10) Near-Earth ICME speed measured by \textit{in-situ} observation at 1 AU; 
(11)\,--\,(12); Index $\alpha$ and coefficient $\beta$ for a power-law form of radial speed evolution; (13) 1 AU transit speed derived from the CME appearance and the ICME detection times; 
(14)\,--\,(15) Average and standard errors for the speed of the background wind $V_\mathrm{bg}$ measured by spacecraft.
} & ~\\
      \end{tabular}
      }
      \end{table}
    \end{landscape}

\begin{figure}
         \vspace{0.025 \textwidth}
          \centerline{\small \bf     
                       \hspace{0.10 \textwidth}  \color{black}{(a)}
                       \hspace{0.432 \textwidth}  \color{black}{(b)}
                       \hfill
                       }
           \vspace{-0.050 \textwidth}
         \centerline{\includegraphics[width=0.46 \textwidth,clip=]{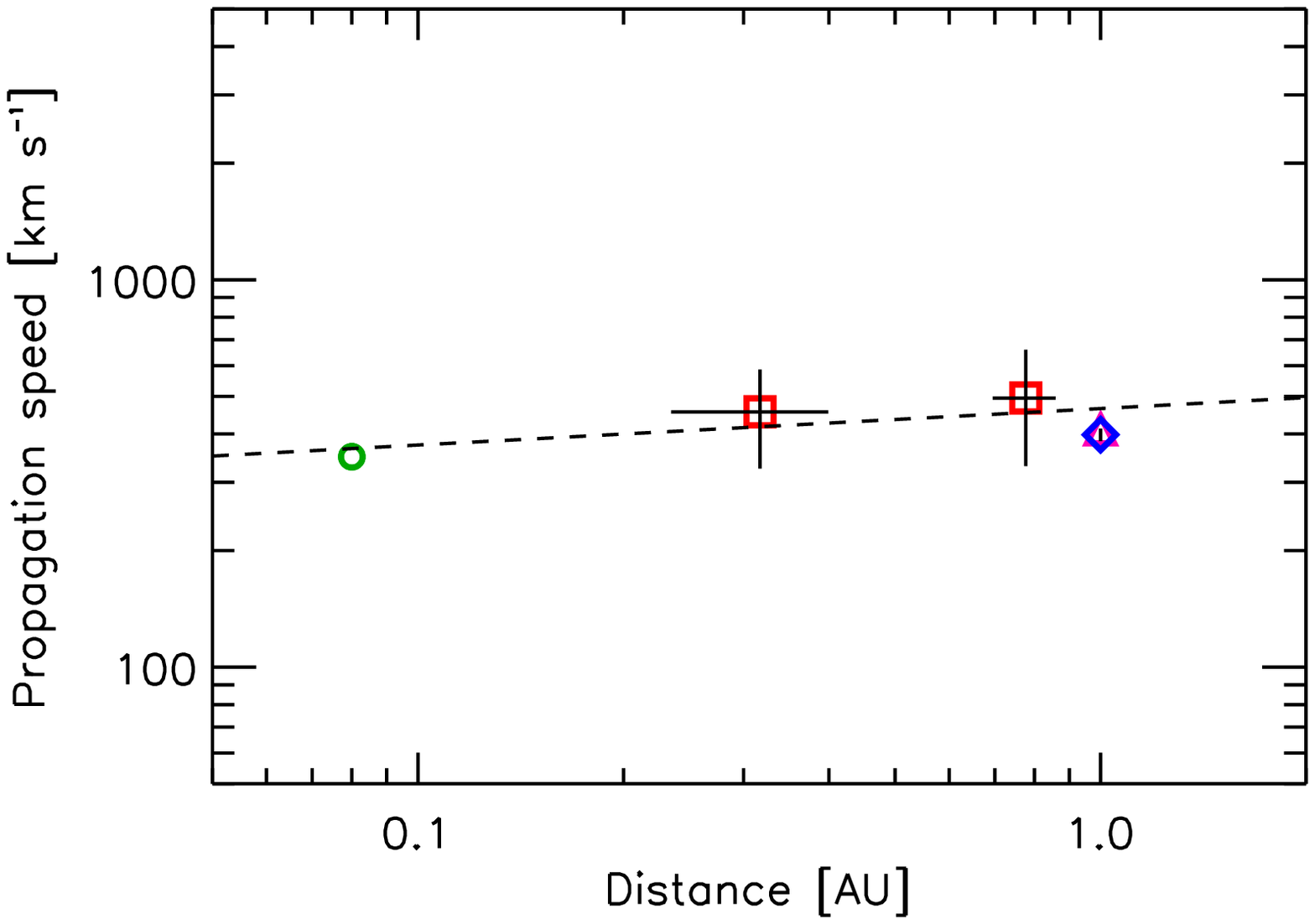}
                     \hspace*{0.01 \textwidth}
                     \includegraphics[width=0.46 \textwidth,clip=]{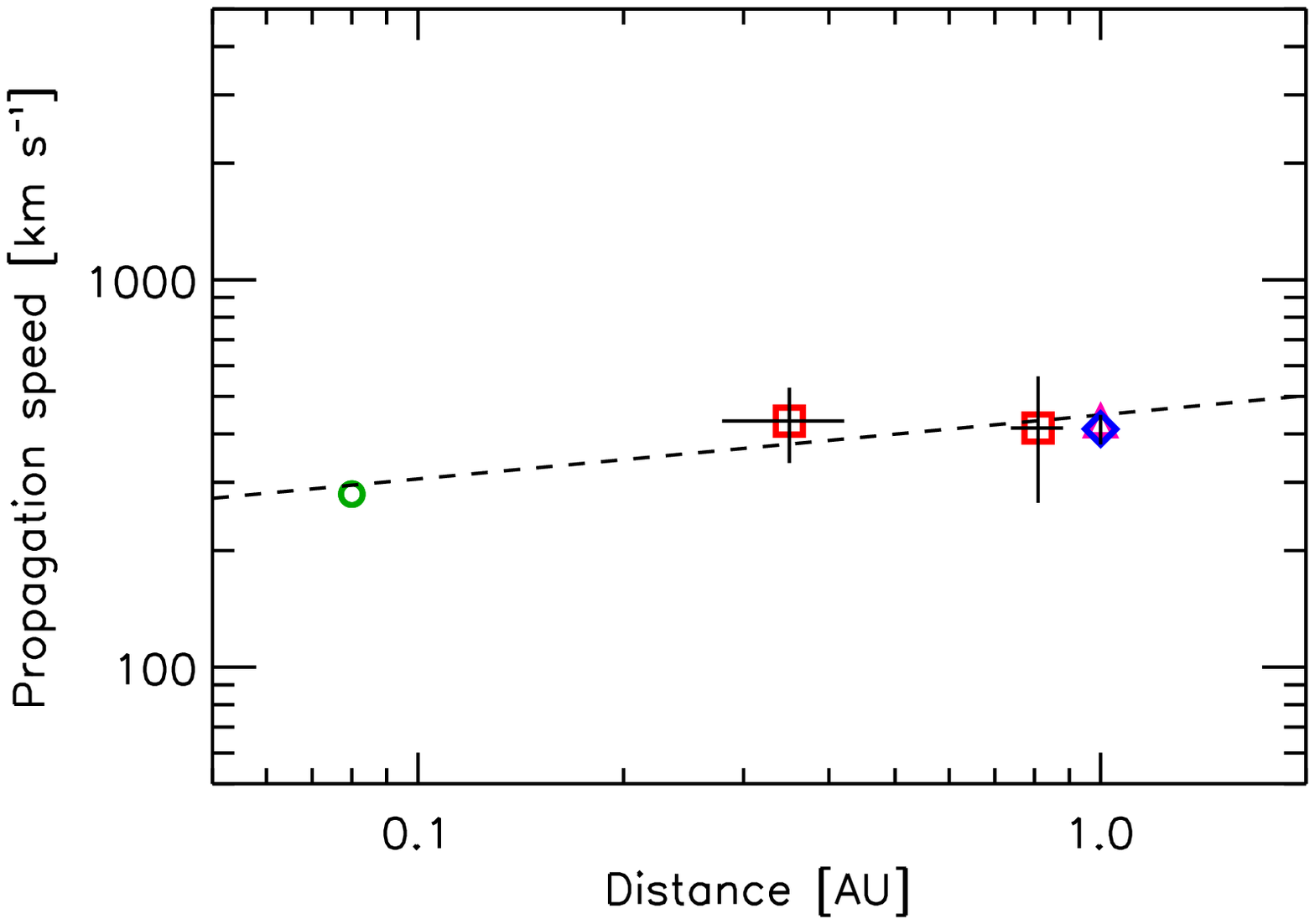}
                     }
           \vspace{0.040 \textwidth} 
         \centerline{\small \bf     
                       \hspace{0.10 \textwidth}  \color{black}{(c)}
                       \hspace{0.432 \textwidth}  \color{black}{(d)}
                       \hfill
                       }
           \vspace{-0.050 \textwidth}
         \centerline{\includegraphics[width=0.46 \textwidth,clip=]{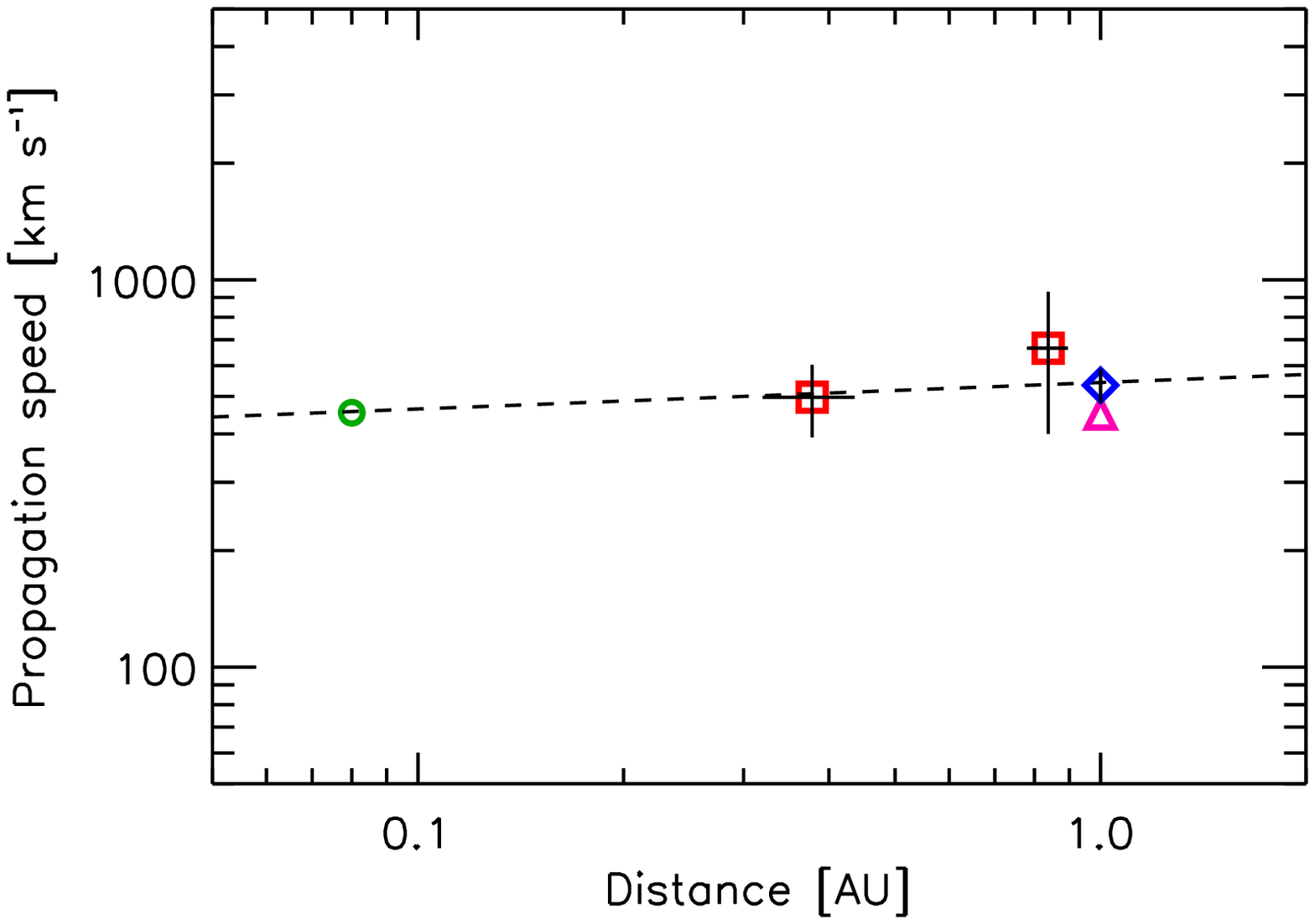}
                     \hspace*{0.01 \textwidth}
                     \includegraphics[width=0.46 \textwidth,clip=]{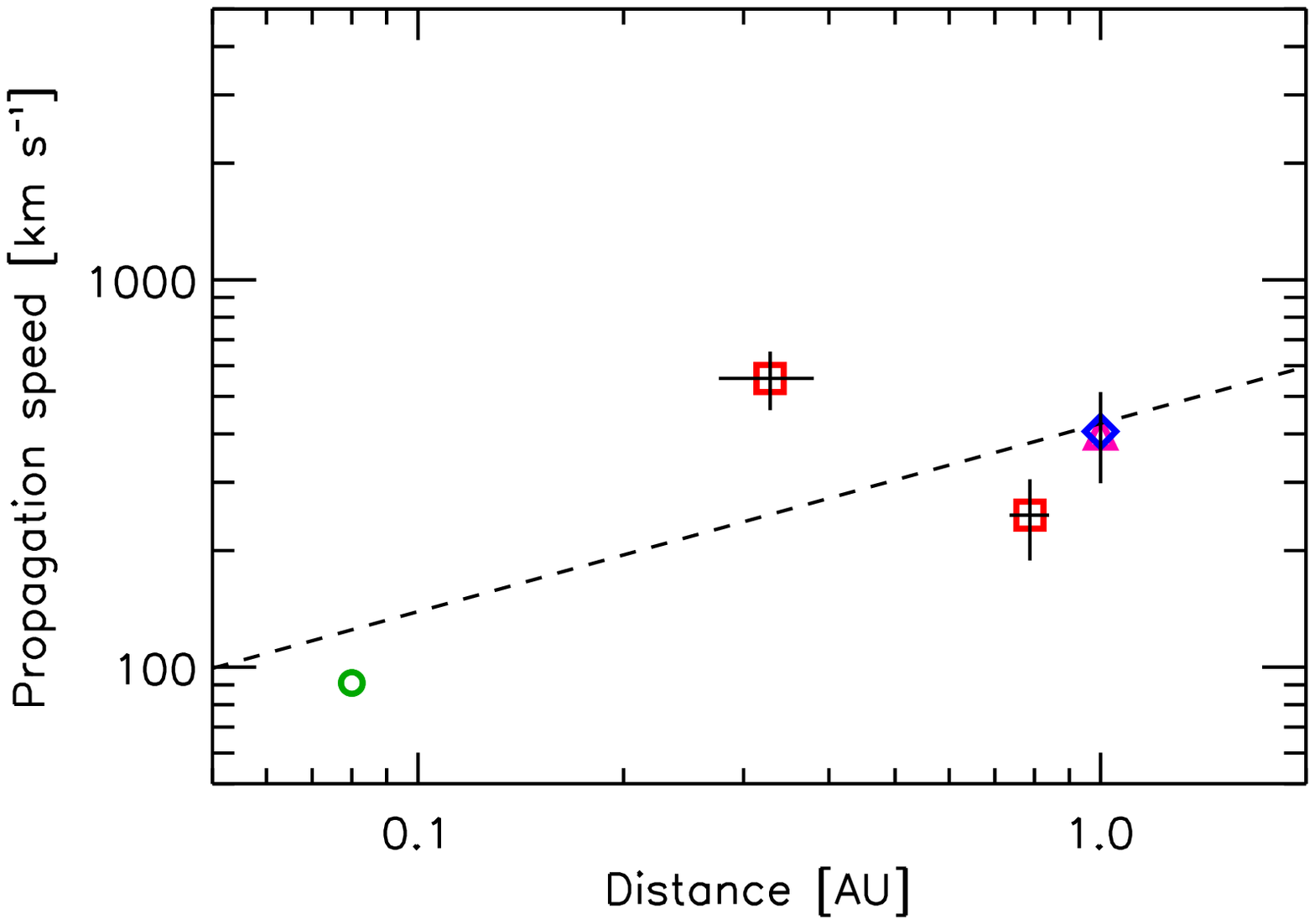}
                     }
           \vspace{0.040 \textwidth} 
         \centerline{\small \bf     
                       \hspace{0.10 \textwidth}  \color{black}{(e)}
                       \hspace{0.432 \textwidth}  \color{black}{(f)}
                       \hfill
                       }
           \vspace{-0.050 \textwidth}
         \centerline{\includegraphics[width=0.46 \textwidth,clip=]{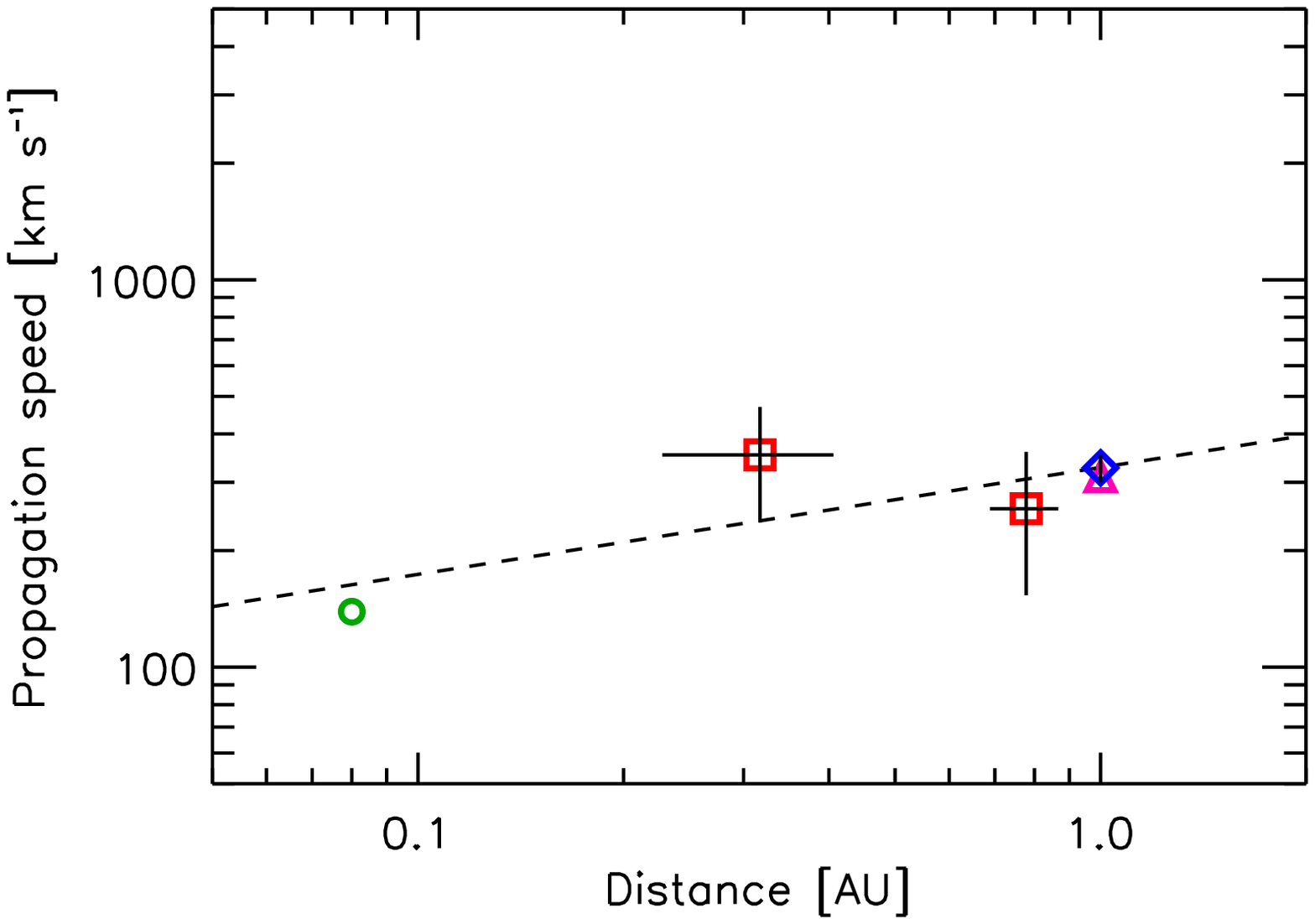}
                     \hspace*{0.01 \textwidth}
                     \includegraphics[width=0.46 \textwidth,clip=]{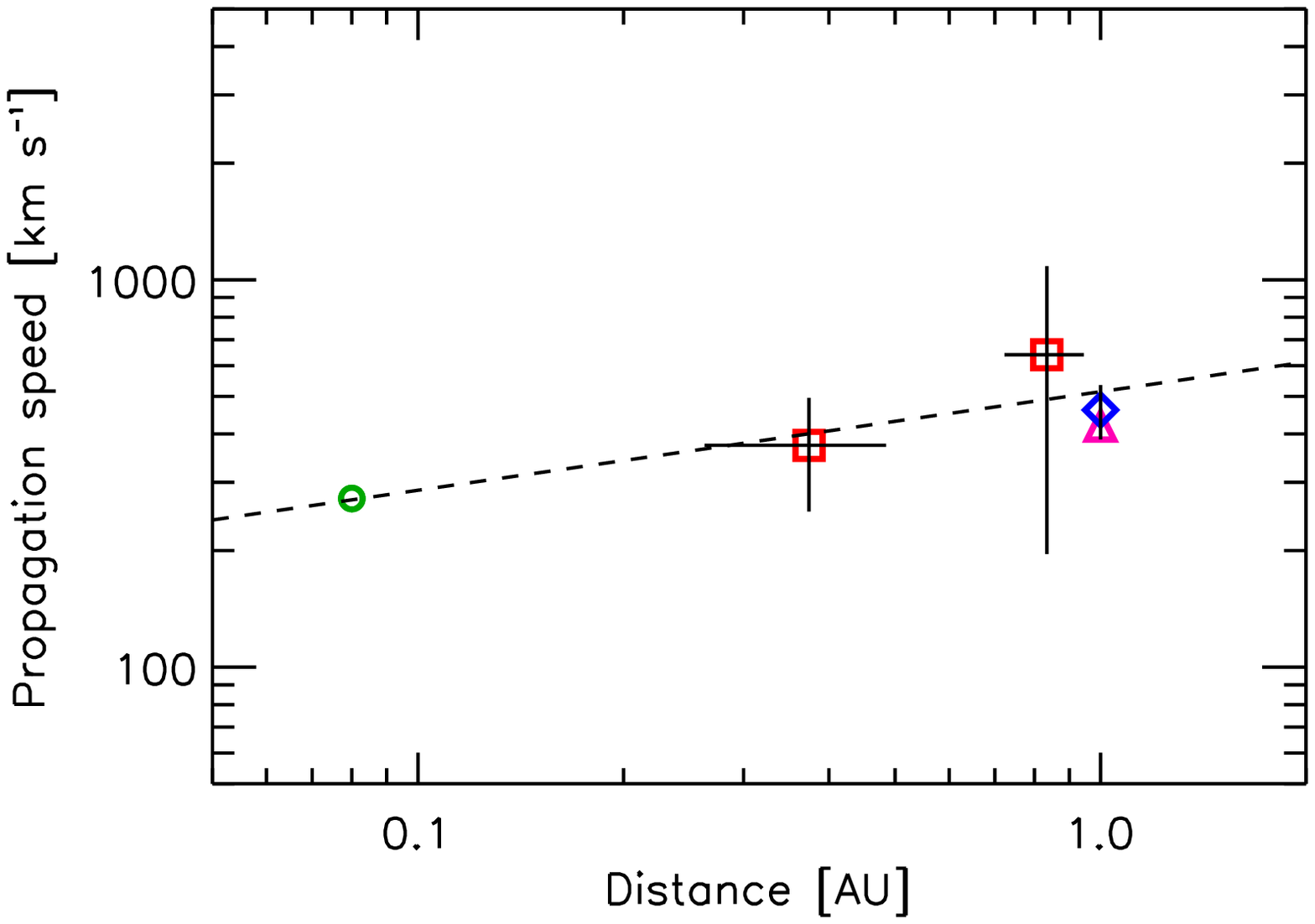}
                     }
           \vspace{0.015 \textwidth} 
         \caption{
         Speed profiles for six slow ICMEs detected between (a) 13 and 16 April 1999, 
         (b) 6 and 10 August 2000, (c) 14 and 18 August 2003, (d) 12 and 17 September 2008,
         (e)29 May and 4 June 2009, and (f) 8 and 12 April 2010. In each panel, the circle (green, at 0.08 AU), squares 
         (red, at ${R_\mathrm{1}}$ and ${R_\mathrm{2}}$), and triangle (purple, at 1 AU) denote measurements of ICME 
         speeds from SOHO/LASCO, IPS, and \textit{in-situ} observations, respectively. 
         Diamonds (blue, at 1 AU) indicate the speed of the background solar wind measured by \textit{in-situ} 
         observations, and the dashed line represents the power-law fit to the data using Equation (\ref{eq.powerlaw}).
          }
         \label{fig1}
         \end{figure}

In the following figures, an error bar represents one standard deviation ($1 \sigma$ error) of the mean for each parameter. 
Figure \ref{fig1} shows speed profiles for six slow ICMEs identified in this study. 
As shown here, the ICME speeds increase with the radial distance, and those at 1 AU are close to the speed of 
the solar wind for all of them. From this, we claim that their speed profiles are well fit by a power-law function within the error bars, 
excluding the 12 September 2008 event (see No.4 in Tables 3 and 4). The 12 September 2008 event has the largest difference in speed 
($V_\mathrm{SOHO} - V_\mathrm{bg} = -314$ ${\mathrm{km~s^{-1}}}$) in our sample, while the others have 
$V_\mathrm{SOHO} - V_\mathrm{bg} > -200$ ${\mathrm{km~s^{-1}}}$. 

\begin{figure}
      \begin{center}
      \centerline{\includegraphics[width=0.9\textwidth,clip=]{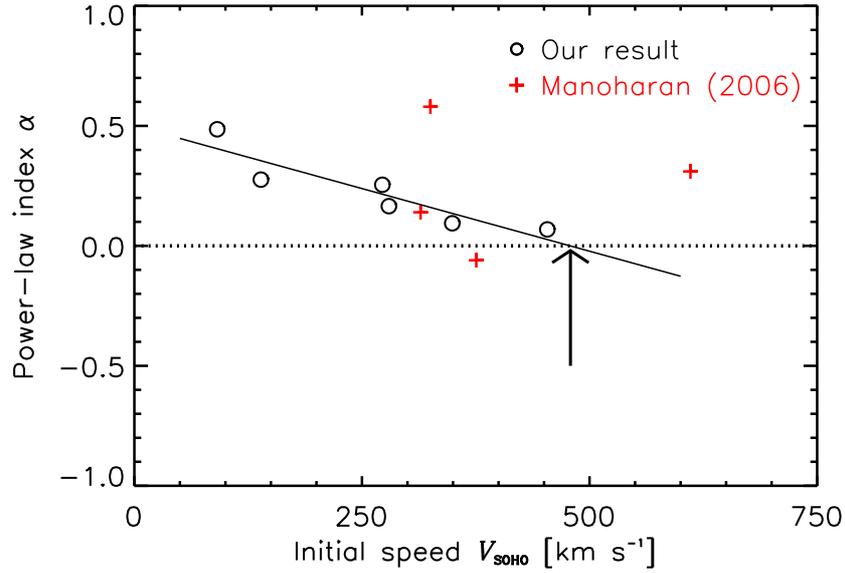}}
      \vspace{-0.05 \textwidth}
      \caption{
      Relationship between the estimated initial speed ${V_\mathrm{SOHO}}$ and the index ${\alpha}$ [Equation (\ref{eq.powerlaw})] for six slow ICMEs. 
      Circles show our data points, and crosses indicate those for the four slow events studied by Manoharan (2006). 
      The solid and dotted lines denote the best-fit line $\alpha = k_\mathrm{1} + k_\mathrm{2}V_\mathrm{SOHO}$ and the $\alpha = 0$ line, respectively. 
      The arrow indicates the intersection of these two lines corresponding to the zero-acceleration point, $V_\mathrm{c} = 479 \pm 126$ 
      ${\mathrm{km~s^{-1}}}$.
      }
      \label{fig2}
      \end{center}
      \end{figure}

Figure \ref{fig2} exhibits the relationship between the initial speed 
$V_\mathrm{SOHO}$ and the index ${\alpha}$ for slow ICMEs. As shown here, ${\alpha}$ decreases from $0.486$ to $0.068$ as 
${V_\mathrm{SOHO}}$ increases. The intersection point between the best-fit line $\alpha = k_\mathrm{1} + k_\mathrm{2}V_\mathrm{SOHO}$ and 
the ${\alpha = 0}$ line is designated as ${V_\mathrm{c}}$ in the following. \inlinecite{Manoharan2006} studied the same relationship 
for 30 CMEs. Four of them are slow CMEs, which have a slower initial speed than the final. 
In order to compare with our result, they are also plotted in Figure \ref{fig2}. We find that their values of ${\alpha}$ range from $0.58$ to $-0.06$. 
The mean values of ${V_\mathrm{c}}$ and coefficients ${k_\mathrm{1}}$ and ${k_\mathrm{2}}$ for the best-fit line, and their standard ($1 \sigma$) errors are given 
in Table \ref{tablebestfit}. From the above examination, we find $V_\mathrm{c} = 479 \pm 126$ ${\mathrm{km~s^{-1}}}$ as the threshold speed 
when ${\alpha}$ becomes zero, \textit{i.e.} the slow ICMEs have zero acceleration.

\begin{figure}
      \leftline{\small \bf     
              \hspace{0.22 \textwidth} \color{black}{(a)}
              \hfill
              }
        \vspace{-0.052 \textwidth}
      \centerline{\includegraphics[width=0.87\textwidth,clip=]{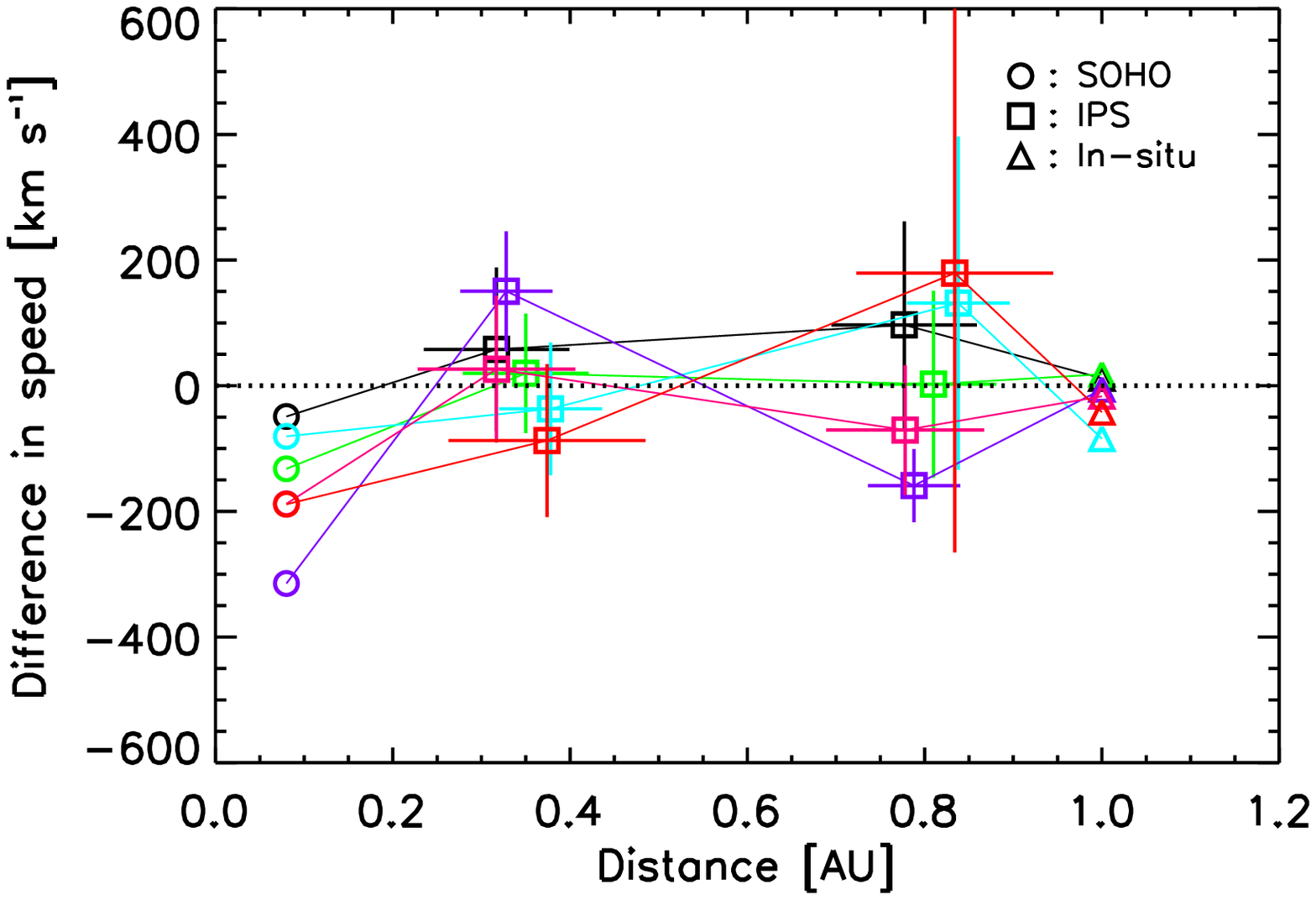}}
      \vspace{0.045\textwidth}
      \leftline{\small \bf     
              \hspace{0.21 \textwidth} \color{black}{(b)}
              \hfill
              }
        \vspace{-0.050 \textwidth}
      \centerline{\includegraphics[width=0.91\textwidth,clip=]{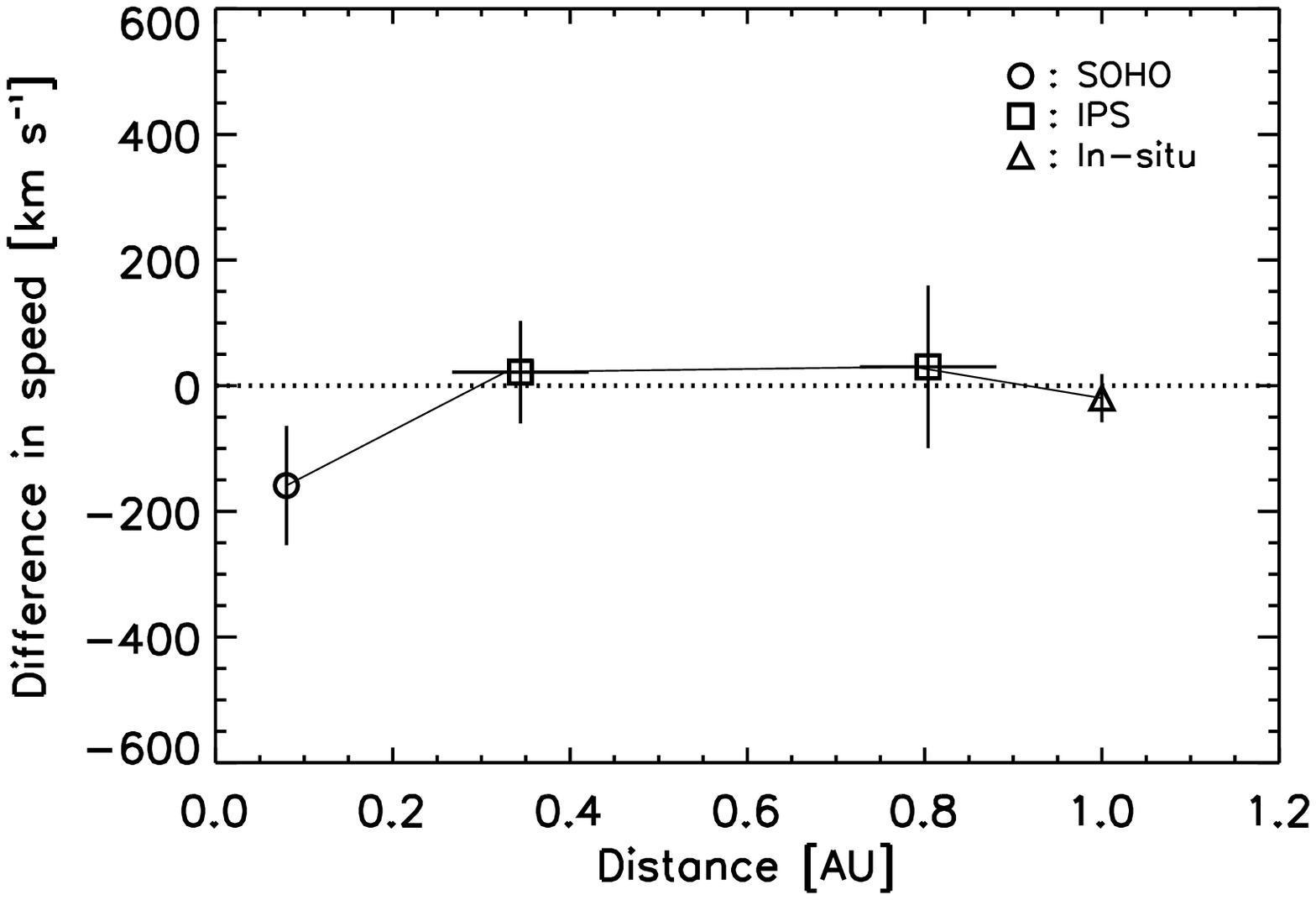}}
      \vspace{0.03\textwidth}
      \caption{
      Radial evolution of (a) differences in speed ($V - V_\mathrm{bg}$) for six slow ICMEs and (b) their averaged profiles. 
      Circles, squares, and triangles indicate the values of ${V - V_\mathrm{bg}}$ for the ICMEs in near-Sun, interplanetary space, 
      and near-Earth regions, respectively. Symbols for each ICME in panel (a) are connected by solid lines with the same color. 
      The dotted line denotes the $V - V_\mathrm{bg} = 0$ line in each panel. 
      }
      \label{fig3}
      \end{figure}

In Figure \ref{fig3}, we plot all of their speed profiles in order to compare radial-speed evolutions of slow ICMEs. 
Here, data points for each ICME are connected by solid lines instead of fitting by 
Equation (\ref{eq.powerlaw}). We note that differences in speed (${V - V_\mathrm{bg}}$) are used instead of ICME speeds in the $y$-axis, 
which correspond to $V_{\mathrm{SOHO}} - V_\mathrm{bg}$ in the near-Sun region, $V_\mathrm{1} - V_\mathrm{bg}$ in the SOHO--IPS region, 
$V_\mathrm{2} - V_\mathrm{bg}$ in the IPS--Earth region, and $V_{\mathrm{Earth}} - V_\mathrm{bg}$ in the near-Earth region. 
From the top panel, we find that the speed differences range from ${-314}$ ${\mathrm{km~s^{-1}}}$ to ${-49}$ ${\mathrm{km~s^{-1}}}$ 
in the near-Sun region, while they show a narrow range from ${-84}$ ${\mathrm{km~s^{-1}}}$ to ${11}$ ${\mathrm{km~s^{-1}}}$ in the near-Earth region. 
The bottom panel shows the averaged profile for their propagation.
Table \ref{tablespeedevolution} gives the average values of the distance and of the speed difference with the standard error in each region for the slow ICMEs.

We attempt to show which  of Equations (\ref{eq.linear}) and (\ref{eq.quadratic}) is more suitable to describe the relationship between 
the acceleration and the difference in speeds for slow ICMEs. 
We assume that ${{\gamma}_\mathrm{1}}$ and ${{\gamma}_\mathrm{2}}$ 
are constants because we want as few variables as possible to describe the relationship as Paper I. 
        %
      %
      %
      %
      \begin{figure}
      \vspace{0.04 \textwidth}
      \leftline{\small \bf     
              \hspace{0.20 \textwidth} \color{black}{(a)}
              \hfill
              }
        \vspace{-0.052 \textwidth}
      \centerline{\includegraphics[width=0.85\textwidth,clip=]{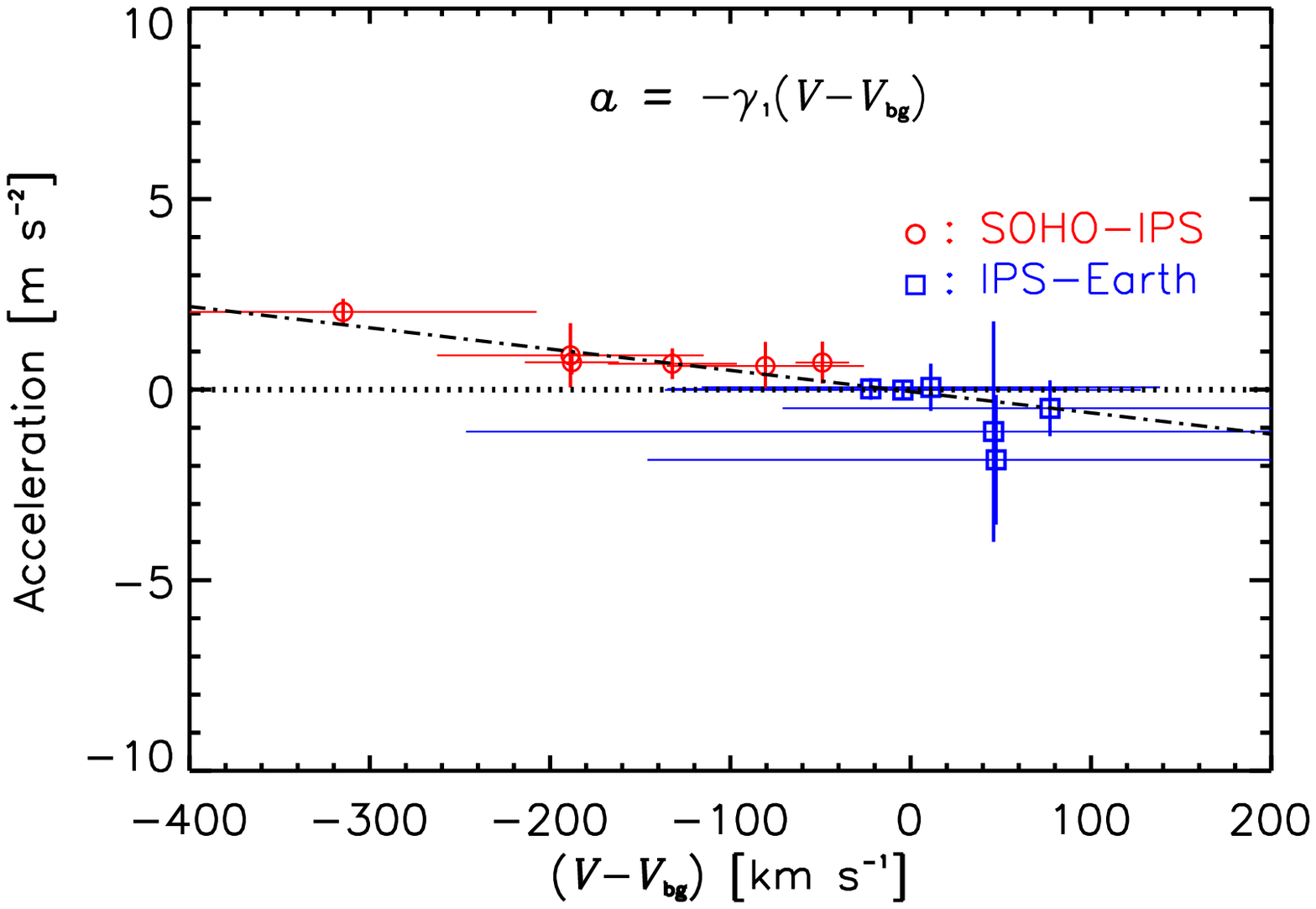}}
      \vspace{0.045\textwidth}
      \leftline{\small \bf     
              \hspace{0.185 \textwidth} \color{black}{(b)}
              \hfill
              }
        \vspace{-0.050 \textwidth}
      \centerline{\includegraphics[width=0.87\textwidth,clip=]{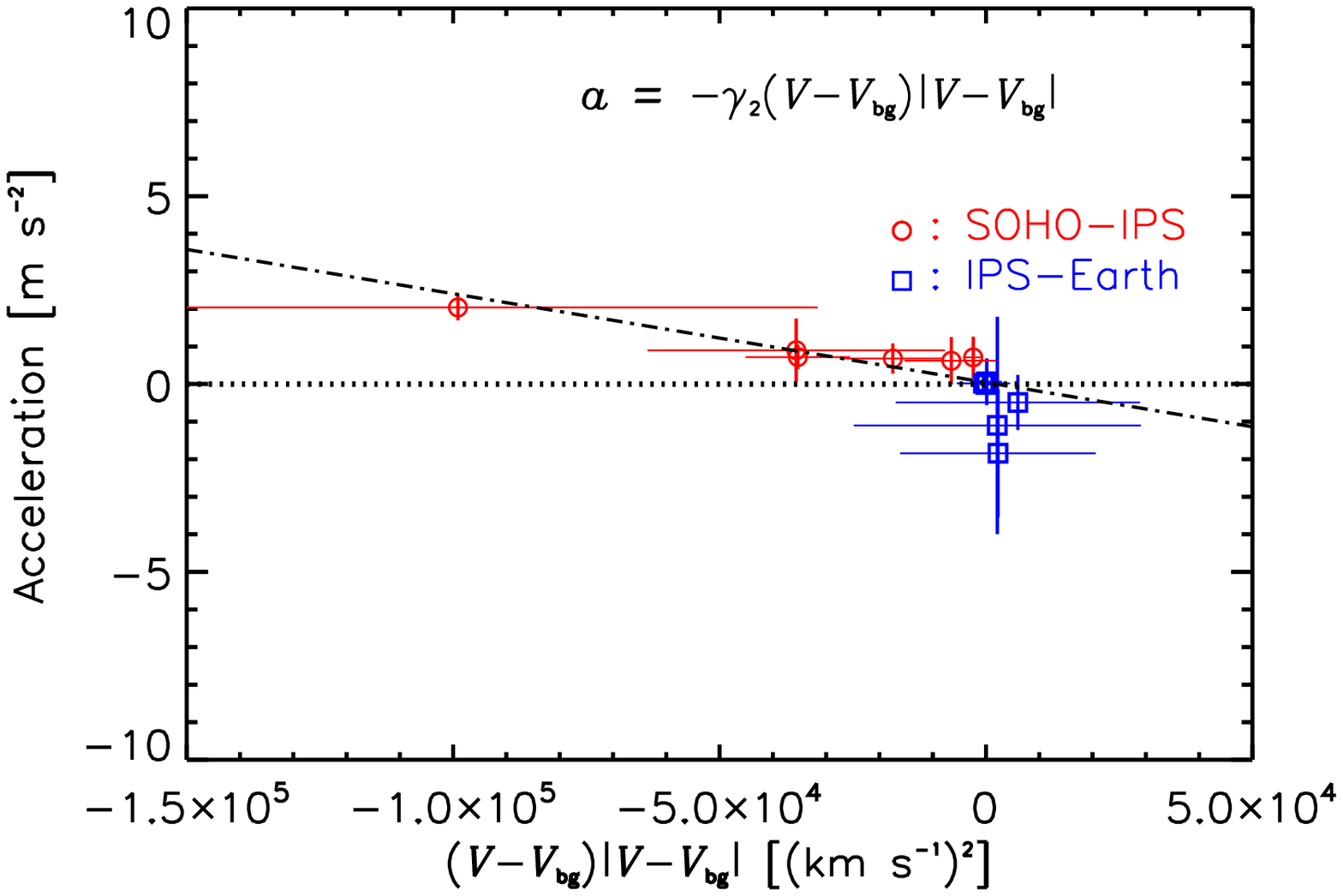}}
      \vspace{0.03\textwidth}
      \caption{
      Relationships between (a) acceleration ($a$) and difference in speed (${V - V_\mathrm{bg}}$) and 
      (b) between $a$ and ${(V - V_\mathrm{bg})|V - V_\mathrm{bg}|}$ for six slow ICMEs in this study. 
      Circles (red) and squares (blue) denote data points in the SOHO--IPS and 
      IPS--Earth regions, respectively. The dash--dotted line and the dotted line denote the best-fit 
      line and the zero-acceleration line, respectively, in each panel.
      }
      \label{fig4}
      \end{figure}
     %
      \begin{figure}
      \vspace{0.04 \textwidth}
      \leftline{\small \bf     
              \hspace{0.21 \textwidth}{(a)}
              \hfill
              }
        \vspace{-0.052 \textwidth}
      \centerline{\includegraphics[width=0.85\textwidth,clip=]{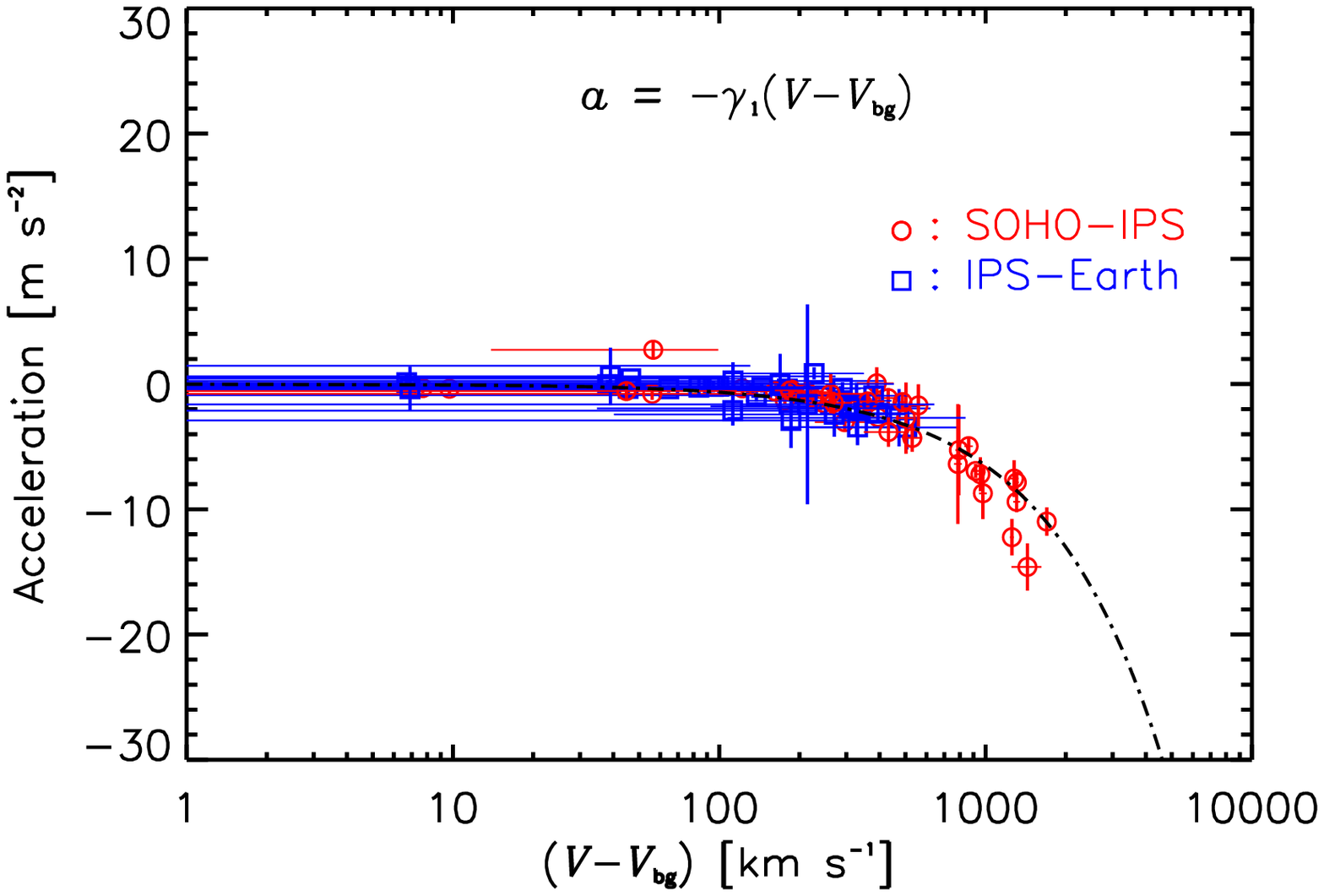}}
      \vspace{0.045\textwidth}
      \leftline{\small \bf     
              \hspace{0.21 \textwidth}{(b)}
              \hfill
              }
        \vspace{-0.050 \textwidth}
      \centerline{\includegraphics[width=0.85\textwidth,clip=]{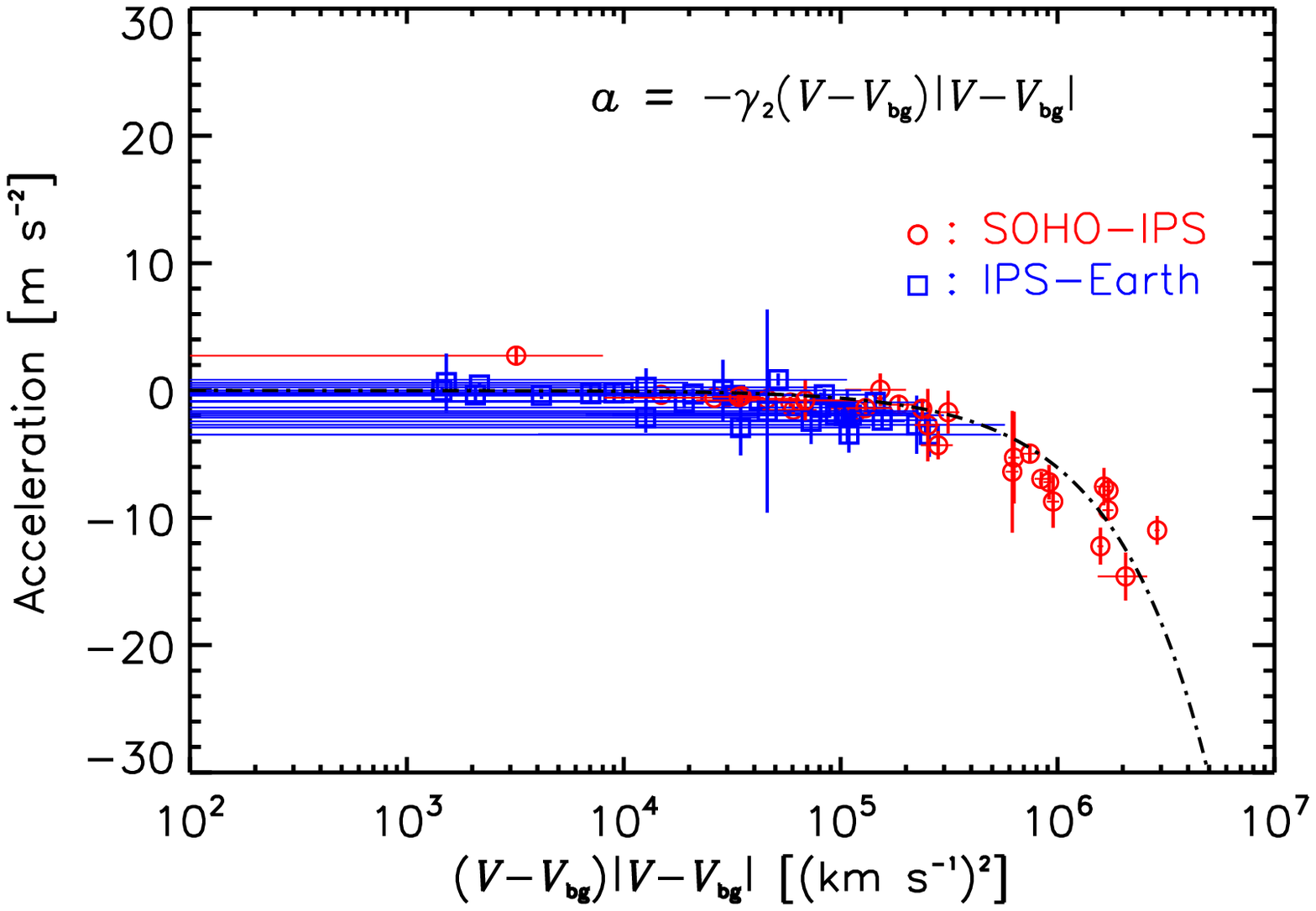}}
      \vspace{0.03\textwidth}
      \caption{
      Relationships between (a) acceleration ($a$) and difference in speed (${V - V_\mathrm{bg}}$) and 
      (b) between $a$ and ${(V - V_\mathrm{bg})|V - V_\mathrm{bg}|}$ for 40 fast and moderate ICMEs 
      (\textit{i.e.} $V_{\mathrm{SOHO}} - V_\mathrm{bg} \ge 0$ ${\mathrm{km~s^{-1}}}$) in this study. 
      Circles (red) and squares (blue) denote data points in the SOHO--IPS and 
      IPS--Earth regions, respectively. The dash--dotted line shows the best-fit 
      line in each panel. 
      }
     \label{fig5}
     \end{figure}
In Figure \ref{fig4}, the top panel shows the relationship between $a$ and $V - V_\mathrm{bg}$, and 
the bottom panel between $a$ and $(V - V_\mathrm{bg})|V - V_\mathrm{bg}|$ for slow ICMEs. 
Data of ${V_{\mathrm{SOHO}}}$ and ${a_\mathrm{1}}$ were used for the SOHO--IPS region, while those of ${V_{\mathrm{IPS}}}$ 
and ${a_\mathrm{2}}$ were used for the IPS--Earth region.
As shown here, the ${\chi}^2$ value for the linear equation is smaller than for the quadratic one. 
In Figure \ref{fig5}, the top panel shows the relationship between $a$ and $V - V_\mathrm{bg}$, and the bottom panel between $a$ and 
$(V - V_\mathrm{bg})|V - V_\mathrm{bg}|$ for a group of fast and moderate ICMEs. 
Data of ${V_{\mathrm{SOHO}}}$ and ${a_\mathrm{1}}$ were used for the SOHO--IPS region, while those of ${V_{\mathrm{IPS}}}$ and 
${a_\mathrm{2}}$ were used for the IPS--Earth region. In this figure, the best fit curves are not straight lines 
because of the logarithmic ${x}$-axis scale. 

We derived the values of coefficients from the slopes of best-fit lines in Figures \ref{fig4} and \ref{fig5}, and also calculated 
the reduced ${\chi}^2$ values in order to assess the goodness of fit. They are listed in Table \ref{tablecoef}. 
     %
      \begin{table}
      \caption{
      Mean values of the distance and of the speed difference with a standard error in each region for the slow ICMEs.
      }
      \label{tablespeedevolution}
      \begin{tabular}{cccc}
      \hline
~~~ & Distance & Difference in speed \\
Region & [AU] & [${\mathrm{km~s^{-1}}}$] \\
      \hline
Near-Sun & ${0.08}$ & ${-159 \pm 95}$ \\
SOHO--IPS & ${0.34 \pm 0.08}$ & ${22 \pm 111}$ \\
IPS--Earth & ${0.80 \pm 0.08}$ & ${30 \pm 197}$ \\
Near-Earth & ${1.00}$ & ${-20 \pm 38}$ \\
      \hline
      \end{tabular}
      \end{table}
     
     %
      \begin{table}
      \caption{
      Mean values of coefficients ${k_\mathrm{1}}$ and ${k_\mathrm{2}}$ for the best-fit line ${\alpha = k_\mathrm{1} + k_\mathrm{2}V_\mathrm{SOHO}}$, 
      the speed at the zero-acceleration point (${V_\mathrm{c}}$), and their standard errors.
      }
     \label{tablebestfit}
     \begin{tabular}{ccccc}
      \hline
~~~ & ${k_\mathrm{1}}$ & ${k_\mathrm{2}}$ & ${V_\mathrm{c}}$~[${\mathrm{km~s^{-1}}}$] \\
      \hline
Mean & ${5.00 \times 10^{-1}}$ & ${-1.04 \times 10^{-3}}$ & 479 \\
Standard error & ${6.82 \times 10^{-2}}$ & ${2.34 \times 10^{-4}}$ & 126 \\
      \hline
      \end{tabular}
      \end{table}
     
     %
      \begin{table}
      \caption{
      Coefficients ${\gamma_\mathrm{1}}$, ${\gamma_\mathrm{2}}$, and reduced $\chi^2$ values for the slow ICMEs and 
      a group of moderate and fast ones.
      }
      \label{tablecoef}
      \begin{tabular}{cccc}
      \hline
Type of ICMEs & Coefficient  & ~~~ \\
and equation & (Mean and standard error) & $\chi^2$ \\
      \hline
\multicolumn{2}{l}{Slow ICMEs} &  ~~~ \\
Linear & ${{\gamma}_{\mathrm{1}} = 5.58 (\pm 1.77) \times 10^{-6}}$~[${\mathrm{s^{-1}}}$] & 0.24 \\
Quadratic & ${{\gamma}_{\mathrm{2}} = 2.36 (\pm 1.03) \times 10^{-11}}$~[${\mathrm{m^{-1}}}$] & 0.36 \\
\multicolumn{2}{l}{Fast and Moderate ICMEs} & ~~~ \\
Linear & ${{\gamma}_{\mathrm{1}} = 6.51 (\pm 0.23) \times 10^{-6}}$~[${\mathrm{s^{-1}}}$] & 1.14 \\
Quadratic & ${{\gamma}_{\mathrm{2}} = 6.06 (\pm 0.23) \times 10^{-12}}$~[${\mathrm{m^{-1}}}$]  & 2.50 \\
      \hline
      \end{tabular}
      \end{table}
     

\section{Discussion} 
      \label{discussion} 

\subsection{Kinematics of Slow ICMEs}
  \label{slowICMEs}

From Figure \ref{fig1}, we confirm that all of the slow ICMEs accelerate toward the speed of the background solar wind during 
their outward propagation. Figure \ref{fig2} shows that the value of ${\alpha}$ decreases from 0.486 to 0.068 as 
the wind speed increases, up to the intersection point ${V_\mathrm{c}}$ where ${\alpha = 0}$. 
Our range of ${\alpha}$ is consistent with that reported by \inlinecite{Manoharan2006} for slow CMEs. 
As presented in Table \ref{tablebestfit}, we derive the coefficients for the best-fit line and the value of ${V_\mathrm{c}}$ 
from the observational data. 
We find that our result ($\alpha =  0.50 - 0.0010V_\mathrm{SOHO}$) is similar to $\alpha =  0.69 - 0.0012V$ 
reported by \inlinecite{Shanmugaraju2009} for the best-fit line. We note that their result was obtained from SOHO/LASCO observations with 
FOV $< 32$ solar radii, while we studied the radial evolution of ICMEs in a wider region from the Sun to the Earth. 
The similarity between these best-fit lines implies that slow ICMEs quickly adjust to the speed of the solar wind. 
We also obtain $V_\mathrm{c} = 479 \pm 126$ ${\mathrm{km~s^{-1}}}$ as the threshold speed where ${\alpha}$ becomes 
zero, which is consistent with Paper I. The mean value is somewhat lower than 
the threshold speed of $575$ ${\mathrm{km~s^{-1}}}$ derived from their best-fit equation, though the difference is within 
the standard error. 

Figure \ref{fig3} (a) shows that the distribution of speed differences in the near-Sun region is wider than 
in the near-Earth region. This and the above results justify our assumption that the motion of ICMEs is controlled by 
the drag force(s) due to interaction with the background solar wind. \inlinecite{Temmer2011}, \inlinecite{Rollett2012}, and 
\inlinecite{Vrsnak2013} reported that the acceleration of slow ICMEs attains the speed of the solar wind within 0.5 AU. 
Figure \ref{fig3} (b) and Table \ref{tablespeedevolution} show that the slow ICMEs attain their final speed by 
$0.34 \pm 0.03$ AU. These are consistent with the earlier studies. It is emphasized that the acceleration cessation distance of 
${\approx 0.3}$ AU for slow ICMEs is different from ${\approx 0.8}$ AU for fast ones as reported by 
\inlinecite{Gopalswamy2001} and Paper I. Using the numerical MHD simulation, \inlinecite{Vrsnak2010} found that 
ICMEs having a large angular width adjust to the speed of the solar wind already close to the Sun. 

We confirm that not only a group of fast and moderate ICMEs, but also slow ICMEs show 
that the ${\chi}^2$ value for the linear equation is smaller than for the quadratic one. 
However, the assessment of significance level shows that Equation (\ref{eq.quadratic}) is more suitable 
than (\ref{eq.linear}) to describe the relationship between $a$ and $V - V_\mathrm{bg}$ for the slow ICMEs because
the latter is too good to fit with data points. 
\inlinecite{Maloney2010} introduced an equation of motion 
${\mathrm{d}}V / {\mathrm{d}}t = -{\kappa}R^{-{\lambda}}(V - V_\mathrm{bg})^{\phi}$ in order to describe the motion of ICMEs, 
where ${\kappa}$, ${\lambda}$, and ${\phi}$ are constants. 
They reported that a quadratic equation ${({\phi} = 2)}$ explained the motion of a slow ICME, 
while a linear equation ${({\phi} = 1)}$ gave a better fitting than the quadratic one for 
the motion of a fast ICME. \inlinecite{Byrne2010} also presented evidence that the aerodynamic drag force acted on 
a slow ICME of 12 December 2008. 
Our results are consistent with their studies for slow ICMEs. 
However, six events of slow ICMEs in our sample are not sufficient to investigate their kinematics more precisely, 
while we detected 40 events of fast and moderate ones during 1997\,--\,2011. 
We need to identify more slow ICMEs and then examine their propagation carefully.

\subsection{Modified Drag Equation for Fast and Moderate ICMEs}
  \label{modified equation}

For the group of fast and moderate ICMEs, we find the values of coefficients ${\gamma}_\mathrm{1}$ 
and ${\gamma}_\mathrm{2}$, which are consistent with Paper I.
Although the constancy of these coefficients is assumed in Equations (\ref{eq.linear}) and (\ref{eq.quadratic}), 
we also find a speed dependence in ${{\gamma}_\mathrm{1}}$ and ${{\gamma}_\mathrm{2}}$ as shown in Table \ref{tablecoef}.
In Paper I, we showed a linear relationship between the acceleration and difference in speed for 
a group of fast and moderate ICMEs, and then proposed a simple expression:
\begin{equation}
    \label{eq.previous}
    a = -6.58 \times 10^{-6}(V - V_\mathrm{bg}),
\end{equation}
as an equation of ICME motion on the assumption that the coefficient is constant 
in a speed range of $V_{\mathrm{SOHO}} - V_\mathrm{bg} \ge 0~{\mathrm{km~s^{-1}}}$. Now, we need to correct our assumption 
for the constancy of ${{\gamma}_\mathrm{1}}$. 

In order to analyze this point in detail, we calculated 
the mean values of ${{\gamma}_\mathrm{1}}$ and difference in speed with the standard ($1 \sigma$) error for each classification of ICMEs. 
These values are presented in Table \ref{table8}. In this analysis, slow ICMEs are excluded from consideration because of 
the conclusion presented in the previous subsection. Earlier studies (\textit{e.g.} \opencite{Vrsnak2002}; 
\opencite{Maloney2010}) assumed a distance dependence of ${{\gamma}_\mathrm{1}}$ such as ${{\kappa}R^{-{\lambda}}}$. 
We also examined the difference between ${{\gamma}_\mathrm{1}}$ in the SOHO--IPS region and in the IPS--Earth region for fast and 
moderate ICMEs. The mean values of ${{\gamma}_\mathrm{1}}$ and the distance, with the standard error in each region, are given in Table \ref{table9}. 
From comparison between the above results, we find that a speed dependence of ${{\gamma}_\mathrm{1}}$ is more 
significant than its distance dependence. Therefore, we conclude that the former is a more remarkable factor than the latter 
in the following examination. We used the values of mean difference in speed and ${{\gamma}_\mathrm{1}}$ for fast and moderate 
ICMEs, and draw the straight line through their data points on a $x$-$y$-chart. 
From the mean values of the slope and of the intercept in the $y$-axis, 
the relationship between ${{\gamma}_\mathrm{1}}$ and mean $V - V_\mathrm{bg}$ can be approximated by 
the following equation:
\begin{equation}
   \label{eq.gamma}
   {\gamma}_{1} = 2.07 \times 10^{-12}(V - V_\mathrm{bg}) + 4.84 \times 10^{-6}.
\end{equation}
We modify the expression for the ICME motion by taking 
the variability of ${{\gamma}_\mathrm{1}}$ into account. Substituting Equation (\ref{eq.gamma}) into 
Equation (\ref{eq.linear}), we obtain the following expression:
\begin{equation}
      \label{eq.modified}
      a = -2.07 \times 10^{-12}(V - V_\mathrm{bg})|V - V_\mathrm{bg}| -4.84 \times 10^{-6}(V - V_\mathrm{bg}).
\end{equation}

\begin{figure}
      \begin{center}
      \centerline{\includegraphics[width=0.9\textwidth,clip=]{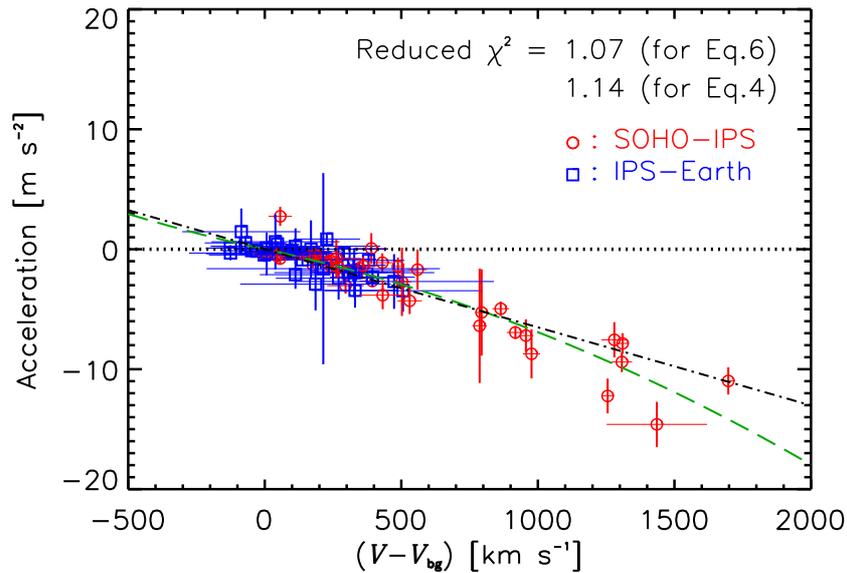}}
      \vspace{-0.05 \textwidth}
      \caption{
      Relationship between acceleration $a$ and difference in speed ${(V - V_\mathrm{bg})}$ for 40 fast and moderate ICMEs 
      identified in this study. 
      Circles (red) and squares (blue) denote data points in the SOHO--IPS and 
      IPS--Earth regions, respectively. 
      The dotted line shows the zero-acceleration line. The dash--dotted (black) line and the dashed (green) line denote 
      the acceleration-speed profiles of Equation (\ref{eq.previous}) and Equation (\ref{eq.modified}), 
      respectively. 
      }
      \label{fig6}
      \end{center}
      \end{figure}

Acceleration-speed profiles given by Equations (\ref{eq.previous}) and (\ref{eq.modified}) 
were compared with observations in Figure \ref{fig6}. 
Data of ${V_{\mathrm{SOHO}}}$ and ${a_\mathrm{1}}$ were used for the SOHO--IPS region, while those of 
${V_{\mathrm{IPS}}}$ and ${a_{2}}$ were used for the IPS--Earth region. 
We confirm that the ${\chi}^2$ value for Equation (\ref{eq.modified}) is more closer 
to unity than that for Equation (\ref{eq.previous}), although both ${\chi}^2$ values satisfy the statistical significance
level of 0.05. Therefore, we conclude that Equation (\ref{eq.modified}) is more appropriate than Equation (\ref{eq.previous}) to 
describe the motion of ICMEs propagating faster than the solar wind. 
On the other hand, we also confirm that the acceleration-speed profile given by Equation (\ref{eq.modified})
is very close to that of Equation (\ref{eq.previous}) with a discrepancy of $< \pm 0.4~\mathrm{m~s^{-2}}$ in a range of 
speed from $0$ to ${\approx 1000~{\mathrm{km~s^{-1}}}}$.
This confirmation suggests that Equation (\ref{eq.previous}) is a good approximation for kinematics of ICMEs with 
${0}$ ${{\mathrm{km~s^{-1}}}}$ $\le V - V_\mathrm{bg} < 1000$ ${\mathrm{km~s^{-1}}}$.
     %
      \begin{table}
      \caption{
      Mean values of coefficient ${\gamma_\mathrm{1}}$ and the difference in speed with the standard errors for each group of ICMEs.
      }
      \label{table8}
      \begin{tabular}{ccccc}
      \hline
Type of& ${V_{\mathrm{SOHO}} - V_\mathrm{bg}}$ & ${\gamma_\mathrm{1}}$ \\
ICMEs & [${\mathrm{km~s^{-1}}}$] & [${\times 10^{-6}}$ ${\mathrm{s^{-1}}}$] \\
      \hline
Fast & ~$1012 \pm 357$ & ${6.94 \pm 0.26}$ \\
Moderate & ~$~231 \pm 138$ & ${5.32 \pm 0.40}$ \\
     \hline
      \end{tabular}
      \end{table}
     %
      \begin{table}
      \caption{
      Mean values of coefficient ${\gamma_\mathrm{1}}$ and the distance in the SOHO--IPS and IPS--Earth regions 
      for each group of ICMEs, and their standard errors.
      }
      \label{table9}
      \begin{tabular}{ccccc}
      \hline
~~~ & Distance & \multicolumn{2}{c}{${\gamma_\mathrm{1}}$ [${\times 10^{-6}}$ ${\mathrm{s^{-1}}}$]} \\
Region & [AU] &  Fast ICMEs & Moderate ICMEs \\
      \hline
SOHO--IPS &  ${0.34 \pm 0.08}$  & ${6.95 \pm 0.27}$ & ${5.36 \pm 0.41}$ \\
IPS--Earth &  ${0.80 \pm 0.08}$  & ${6.20 \pm 1.88}$ & ${4.74 \pm 1.66}$ \\
     \hline
      \end{tabular}
      \end{table}
      %

Equations (\ref{eq.linear}) and (\ref{eq.gamma}) are similar to a set of simultaneous equations 
in the ``snow plough'' model proposed by \inlinecite{Tappin2006}. While he explained that 
ICMEs decelerate by the momentum transfer with piling up mass in front of them, we will explain 
their acceleration in terms of fluid dynamics. When an object propagates in a fluid, 
the object suffers the drag force due to the interaction with the surrounding medium. 
The characteristics of drag force changes depending on the Reynolds number 
$\mathrm{Re} = ({\rho}dU) / {\mu} = (dU) / {\nu}$, 
where ${\rho}$, $d$, $U$, and ${\mu}$ are fluid density, the size and speed of the object, and 
viscosity of the fluid, respectively, and ${\nu}$ is the kinematic viscosity defined as 
${{\nu} = {\mu} / {\rho}}$. 
The drag force is proportional to $U$ for ${\mathrm{Re} \ll 1}$ and to ${U^{2}}$ for ${\mathrm{Re} \gg 1}$. 
The former is called the hydrodynamic Stokes drag force, while the latter corresponds to the aerodynamic drag force. 
\citeauthor{Borgazzi2009} (\citeyear{Borgazzi2008}, \citeyear{Borgazzi2009}) discussed the ICME propagation 
using a model involving both drag forces by assuming a spherical body of ICMEs. 
They assumed that ICMEs undergo both Stokes and aerodynamic drag forces during their propagation. 

Here, we note that Equation (\ref{eq.modified}) consists of quadratic and linear terms, which can be interpreted to be due to 
the aerodynamic and Stokes drag forces, respectively. 
In order to understand the character of this equation, we assess the contribution from each term to 
the net acceleration. Such contributions were calculated by applying various solar-wind speeds to Equation (\ref{eq.modified}), 
and are listed in Table \ref{table10}. 
We confirm from this table that the contribution from each term varies with the difference in speed, 
and the linear term (Stokes drag force) has a larger contribution than the quadratic term (aerodynamic drag force) to the net acceleration
in a speed range of ${0}$ ${{\mathrm{km~s^{-1}}}}$ $\le V - V_\mathrm{bg} \le 2300$ ${\mathrm{km~s^{-1}}}$. This result suggests that 
ICMEs with the above speed range are controlled mainly by 
the Stokes drag force, while the aerodynamic drag force is a predominant factor for 
the propagation of ICMEs with ${V - V_\mathrm{bg} > 2300}$ ${\mathrm{km~s^{-1}}}$. This interpretation is consistent with 
the fundamental theory of fluid dynamics because ${\mathrm{Re} \propto V}$. However, ICMEs with $V - V_\mathrm{bg} > 2300$ 
${\mathrm{km~s^{-1}}}$ are extremely fast eruptions having the propagation speed exceeding 2800 ${\mathrm{km~s^{-1}}}$, and so are very 
rare \cite{Gopalswamy2009}. Therefore, we conclude that the Stokes drag force will play the key role for almost all of the fast and moderate ICMEs. 
     %
      \begin{table}
      \caption{
      Contributions of linear and quadratic terms to the net acceleration versus the difference 
      in speed (${V - V_\mathrm{bg}}$) in Equation (\ref{eq.modified}).
      }
      \label{table10}
      \begin{tabular}{ccccc}
      \hline
${(V - V_\mathrm{bg})}$ & Net acceleration & Linear term & Quadratic term \\
~[${\mathrm{km~s^{-1}}}$] & [${\mathrm{m~s^{-2}}}$] & contrib. [\%] & contrib. [\%] \\
      \hline
~100 & ~$-0.51$ & 95.9 & ~4.1 \\
~500 & ~$-2.94$ & 82.4 & 17.6 \\
1000 & ~$-6.91$ & 70.1 & 29.9 \\
1500 & $-11.92$ & 60.9 & 39.1 \\
2000 & $-17.97$ & 53.9 & 46.1 \\
2100 & $-19.30$ & 52.7 & 47.3 \\
2200 & $-20.68$ & 51.5 & 48.5 \\
2300 & $-22.09$ & 50.4 & 49.6 \\
2400 & $-23.55$ & 49.4 & 50.6 \\
2500 & $-25.05$ & 48.4 & 51.6 \\
3000 & $-33.16$ & 43.8 & 56.2 \\
     \hline
      \end{tabular}
      \end{table}

\subsection{Kinematic Viscosity and Drag Coefficient for the ICME--Solar Wind Interaction}
  \label{viscosity}

Furthermore, Equation (\ref{eq.modified}) also implies that the effective kinematic viscosity of the solar wind 
(${{\nu}_\mathrm{SW}}$) exhibits a large value in the ICME--solar wind interaction system. Now, we estimate 
the value of ${{\nu}_\mathrm{SW}}$. \inlinecite{Borgazzi2008} pointed out that the drag force is 
represented by $F = -6{\pi}{\nu}{\rho}d(V - V_\mathrm{bg})$ for ${\mathrm{Re} \ll 1}$ 
by assuming a spherical body of ICMEs. 
We apply this expression to the linear term in Equation (\ref{eq.modified}), and find 
\begin{equation}
   \label{eq.viscosity}
\frac{6{\pi}{\nu}_{\mathrm{SW}}{\rho}_{\mathrm{SW}}d}{m} = 4.84 \times 10^{-6}~{\mathrm{s^{-1}}},
\end{equation} 
where ${{\rho}_\mathrm{SW}}$, $d$, and $m$ are the solar wind density, radial size, and mass of ICMEs, respectively. 
Substituting the values of $d = 4.49 \times 10^{10}$ m (${= 0.3}$ AU) \cite{Richardson2010}, 
$m = 1.7 \times 10^{12}$ kg \cite{Vourlidas2002}, and ${\rho}_\mathrm{SW} = 1.67 \times 10^{-22}$ ${\mathrm{kg~m^{-3}}}$ 
(in other words, the total mass density of 10 protons per cubic centimeter) in the above equation, 
we obtain ${{\nu}_\mathrm{SW} = 5.8 \times 10^{16}}$ ${\mathrm{m^{2}~s^{-1}}}$. This value is an order of 
magnitude smaller than the viscosity estimated by \inlinecite{Lara2011}. On the other hand, if we use the value of 
viscosity estimated by them, we may estimate the mass of ICMEs instead of ${{\nu}_\mathrm{SW}}$. \inlinecite{Lara2011} reported 
${\nu}_\mathrm{SW} = 1.55 \times 10^{17}$ ${\mathrm{m^{2}~s^{-1}}}$ (from the speed matching method) and 
${2.60 \times 10^{17}}$ ${\mathrm{m^{2}~s^{-1}}}$ (from the time matching method) as the value of viscosity. Substituting this value into 
Equation (\ref{eq.viscosity}) with the above $d$ and ${\rho}_\mathrm{SW}$ values, we obtain $m \approx 10^{13}$ kg. 
This value corresponds to the upper limit of CME mass observed using SOHO/LASCO \cite{Gopalswamy2009}. Therefore, this analysis corroborates 
our expectation in Paper I that ICMEs detected by the IPS observations are probably massive events.

\inlinecite{Borgazzi2008} also showed that the drag force is described by $F = -(\mathrm{C_{d}}A{\rho})/2(V - V_\mathrm{bg})^{2}$ 
for ${\mathrm{Re} \gg 1}$, where $\mathrm{C_{d}}$ and $A$ are the dimensionless drag coefficient and cross section of ICMEs, respectively. 
By applying this expression to the quadratic term in Equation (\ref{eq.modified}) with $A = {\pi}({d/2})^{2}$, we can estimate the value of 
$\mathrm{C_{d}}$. Using the following equation:
\begin{equation}
   \label{eq.coefficient}
\frac{{\mathrm{C_{d}}}{\rho}_{\mathrm{SW}}{\pi}d^{2}}{8m} = 2.07 \times 10^{-12}~{\mathrm{m^{-1}}},
\end{equation} 
and the above values of $m$, $d$, and ${\rho}_\mathrm{SW}$, we find $\mathrm{C_{d}} = 27$. 
This value is almost three orders of magnitude smaller than the estimation by \inlinecite{Lara2011}; they reported 
${\mathrm{C_{d}} = 2.63 \times 10^{3}}$ (from the speed matching method) and ${1.08 \times 10^{4}}$ (from the time matching method). 
\inlinecite{Borgazzi2009} reported that $\mathrm{C_{d}}$ is 
${0.6 \times 10^{5}}$\,--\,${1.6 \times 10^{5}}$ (considering the variation in radius) or ${2 \times 10^{4}}$\,--\,${8 \times 10^{4}}$ 
(considering the density variation) for the turbulent regime. 
On the other hand, \inlinecite{Cargill2004} showed by numerical simulations that $\mathrm{C_{d}}$ varies slowly between the Sun and the Earth, 
and is roughly unity for dense ICMEs. He also showed that when the ICME and solar wind densities are similar, $\mathrm{C_{d}}$ is larger than unity 
(between 3 and 10), but remains approximately constant with the radial distance. 
As shown here, each researcher reports different values of $\mathrm{C_{d}}$, and so it is difficult to determine 
its real value. 

\section{Summary and Conclusions} 
      \label{conclusion} 

We investigated kinematic properties of six slow (${V_{\mathrm{SOHO}} - V_\mathrm{bg} < 0}$ ${\mathrm{km~s^{-1}}}$), 
25 moderate (${0}$ ${\mathrm{km~s^{-1}}}$ ${\le V_{\mathrm{SOHO}} - V_\mathrm{bg} \le 500}$ ${\mathrm{km~s^{-1}}}$), 
and 15 fast ($V_{\mathrm{SOHO}} - V_\mathrm{bg} > 500$ ${\mathrm{km~s^{-1}}}$) ICMEs detected by SOHO/LASCO, 
IPS, and \textit{in-situ} observations during 1997\,--\,2011. 

Our analyses for the slow ICMEs show the following results: 
i) They accelerate toward the speed of the background solar wind during their propagation, and attain their final speed by 
${0.34 \pm 0.03}$ AU. ii) The acceleration ends when they reach ${479 \pm 126}$ ${\mathrm{km~s^{-1}}}$; this is close to 
the typical speed of the solar wind during the period of this study. 
Examinations of the relationship between the difference in speed and the acceleration
and the assessment of significance level for them show that iii) Equation (\ref{eq.quadratic}) 
with ${\gamma}_\mathrm{2} = 2.36~(\pm 1.03) \times 10^{-11}$ $\mathrm{m^{-1}}$ is more suitable 
than Equation (\ref{eq.linear}) to describe the kinematics of slow ICMEs. The result iii) is consistent with earlier 
studies by \inlinecite{Maloney2010} and \inlinecite{Byrne2010}. 
However, six events of slow ICMEs in our sample are not sufficient to investigate their kinematics 
more precisely. Therefore, we need to identify more slow ICMEs and then examine their kinematics carefully. 

We also found from examinations of fast and moderate ICMEs that the value of coefficient ${{\gamma}_\mathrm{1}}$ has speed dependence 
described by Equation (\ref{eq.gamma}). On the basis of these, 
we find a modified equation, $a = -2.07 \times 10^{-12}$$(V - V_\mathrm{bg})|V - V_\mathrm{bg}|$~$-ð4.84 \times 10^{-6}(V - V_\mathrm{bg})$, 
for the ICME motion. We interpret this equation as indicating that ICMEs with 
${0}$ ${\mathrm{km~s^{-1}}}$ $ \le V - V_\mathrm{bg} \le 2300$ ${\mathrm{km~s^{-1}}}$ are controlled mainly by the Stokes drag force, 
while the aerodynamic drag force is a predominant factor for the propagation of ICMEs 
with $V - V_\mathrm{bg} > 2300$ ${\mathrm{km~s^{-1}}}$. 
Because such extremely fast ICMEs are very rare, we conclude that the Stokes drag force will play the key role for almost all of 
the fast and moderate ICMEs. 

We also estimated the effective kinematic viscosity of the solar wind 
(${{\nu}_\mathrm{SW}}$) and the dimensionless drag coefficient ($\mathrm{C_{d}}$) in the ICME--solar wind interaction system. 
Combining the linear term in the modified equation and $F = -6{\pi}{\nu}{\rho}d(V - V_\mathrm{bg})$, we obtain 
${\nu}_\mathrm{SW} = 5.8 \times 10^{16}$ ${\mathrm{m^{2}~s^{-1}}}$; 
this is an order of magnitude smaller than the value in an earlier study by \inlinecite{Lara2011}. 
By comparing the quadratic term in the modified equation with $F = -(\mathrm{C_{d}}A{\rho})/2(V - V_\mathrm{bg})^{2}$, 
we find $\mathrm{C_{d}} = 27$ for the value of drag coefficient.

%

%
\begin{acks}
The IPS observations were carried out under the solar wind program of 
the Solar-Terrestrial Environment Laboratory (STEL) of Nagoya University. 
We acknowledge use of the SOHO/LASCO CME catalog; this CME catalog is generated and 
maintained at the CDAW Data Center by the National Aeronautics and Space Administration (NASA) 
and the Catholic University of America in cooperation with the Naval Research Laboratory. SOHO is a project of 
international cooperation between the European Space Agency and NASA. We thank the Space Physics Data Facility of 
NASA's Goddard Space Flight Center for use of OMNIWeb service and OMNI data. 
\end{acks}

%
%
 \bibliographystyle{spr-mp-sola}
 \bibliography{slowicme}  

\end{article} 
\end{document}